\documentclass[12pt]{article}
\usepackage{enumerate}
\usepackage{amsfonts}
\usepackage{amsmath}
\usepackage{amssymb}
\usepackage{amsthm}

\def\H{\mathcal{H}}

\def\S{\mathfrak{S}}

\def\T{\mathfrak{T}}
\def\B{\mathfrak{B}}

\newcommand{\rank}{\mathrm{rank}}
\newcommand{\id}{\mathrm{Id}}
\newcommand{\Tr}{\mathrm{Tr}}

\newcommand{\shs}{\hspace{1pt}}

\newcounter{defin}  \newcounter{lemma}  \newcounter{theorem}
\newcounter{property} \newcounter{corol}  \newcounter{remark} \newcounter{example}

\newenvironment{lemma}{\par\refstepcounter{lemma}%\noindent
     \textbf{Lemma \thelemma.} }{\rm\par}

\newenvironment{property}{\par\refstepcounter{property}%\noindent
     \textbf{Proposition \theproperty.}\ }{\rm\par}
\newenvironment{corollary}{\par\refstepcounter{corol}%\noindent
     \textbf{Corollary \thecorol.} }{\rm\par}
\newenvironment{definition}{\par\refstepcounter{defin}%\noindent
     \textbf{Definition \thedefin.}\ }{\rm\par}
\newenvironment{remark}{\par\refstepcounter{remark}%\noindent
     \textbf{Remark \theremark.}}{\rm\par}

\begin{document}

\title{Squashed entanglement in infinite dimensions}
\author{M.E. Shirokov\footnote{Steklov Mathematical Institute, RAS, Moscow, email:msh@mi.ras.ru}}
\date{}
\maketitle

\begin{abstract}
We analyse two possible definitions of the squashed entanglement in an
infinite-dimensional bipartite system: direct translation
of the finite-dimensional definition and its universal extension.  It is  shown that the both definitions produce the same lower semicontinuous entanglement measure possessing all basis properties of the squashed entanglement on the set of states having at least one
finite marginal entropy. Is also shown that the second definition gives an adequate extension of this measure to the set of all states of infinite-dimensional bipartite system.

A general  condition relating continuity of
the squashed entanglement to continuity  of the quantum mutual information is proved and its corollaries are considered.

Continuity bound for the squashed entanglement under the energy constraint on one subsystem is  obtained by using the tight continuity bound for conditional mutual information (proved in the Appendix by using Winter's technique). It is shown that the same continuity bound is valid for the entanglement of formation. As a result the asymptotic continuity of the both entanglement measures under the energy constraint on one subsystem is proved.
\end{abstract}

\tableofcontents

\section{Introduction}

Entanglement is an essential feature of quantum systems which can be
considered as a special quantum correlation having no classical
analogue. One of the main tasks of quantum information theory
consists in finding appropriate quantitative characteristics of
entanglement in bi- and multipartite quantum systems and in studying their
properties \cite{Eisert,ESP,4H,P&V,V}.

Among the existing entanglement measures in finite-dimensional
bipartite systems, the squashed entanglement (introduced
independently in \cite{C&W} and in \cite{Tucci}) is one of the most interesting
one. It possesses all the basic properties of an entanglement
measure including the additivity \cite{SE-F,C&W}. Mathematically, the
squashed entanglement is interesting due to its definition which
includes the infimum of the conditional mutual information over all
extensions $\omega_{ABE}$ of a given state $\omega_{AB}$ with no
restriction on the dimension of the system $E$. This leads to
particular difficulties in proving continuity of the squashed
entanglement and its faithfulness. Since the squashed entanglement was introduced, a lot of papers devoting to analysis of its properties
appeared, see \cite{SE-F,C,W&K,L&W,L&W-2} and the references therein.

In this paper we try to generalize the squashed entanglement to all states of an infinite-dimensional bipartite system by
using two ways:  direct translation
of the finite-dimensional definition and its universal extension (a construction which produces infinite-dimensional entanglement monotone starting from finite-dimensional one, it is described in Section 3 in general settings).

In Section 4 (after a short overview of properties of the squashed entanglement in finite dimensions) we analyse first a
direct infinite-dimensional definition of the squashed entanglement using
the extended quantum conditional mutual information considered in \cite{CMI}. We show that this definition produces
a function possessing all the basic properties of the
squashed entanglement valid in finite dimensions (with the continuity replaced
by the lower semicontinuity) on the set of states having at least one
finite marginal entropy. The main problem  (remained open) is to show that any
separable state in an infinite-dimensional bipartite system can be
extended to a short Markov chain.  This problem (related to the existence
of countably nondecomposable separable states \cite{HSW}) prevents to prove vanishing of the directly defined version of squashed entanglement on all separable states.

Then we consider the universal extension of squashed entanglement -- a lower semicontinuous function on the set of all bipartite states possessing all basis properties of the squashed entanglement valid in finite dimensions. We prove that this extension coincides with the above direct definition of the squashed entanglement on the set of states having at least one
finite marginal entropy. The global coincidence is conjectured but not proved, it is shown to be equivalent to the global lower semicontinuity of the directly defined version of squashed entanglement.

Continuity properties of (the both versions of) the squashed entanglement are analysed in Section 5. We obtain a general  condition relating continuity of the squashed entanglement to continuity of the quantum mutual information and consider its corollaries. In particular, we prove a weak form of the conjecture that local continuity of the squashed entanglement is preserved by local operations. We also consider several simple continuity conditions which can be used in applications.
\smallskip

Then the continuity bound for the squashed entanglement under the energy constraint on one subsystem is obtained by using the tight continuity bound for conditional mutual information (proved in the Appendix by using Winter's technique \cite{Winter}). It is shown that the same continuity bound is valid for the entanglement of formation.  As a result, the asymptotic continuity of the both entanglement measures under the energy constraint on one subsystem is proved.

\section{Preliminaries}

Let $\mathcal{H}$ be a separable Hilbert space,
$\mathfrak{B}(\mathcal{H})$ the algebra of all bounded operators with the operator norm $\|\cdot\|$ and $\mathfrak{T}( \mathcal{H})$ the
Banach space of all trace-class
operators in $\mathcal{H}$  with the trace norm $\|\!\cdot\!\|_1$. Let $\mathfrak{T}_{+}(\mathcal{H})$ be the
cone of positive operators in $\mathfrak{T}( \mathcal{H})$ and
$\mathfrak{S}(\mathcal{H})$  the set of quantum states (operators
in $\mathfrak{T}_{+}(\mathcal{H})$ with unit trace). Note that $\mathfrak{T}_{+}(\mathcal{H})$ and $\mathfrak{S}(\mathcal{H})$
are complete separable metric space with the metric induced by the trace norm \cite{H-SCI,N&Ch}. Denote by $\mathrm{ext}\shs\S(\H)$ the set of all  extreme points of the convex set $\S(\H)$ called \emph{pure states}.

Trace class operators (not only states) will be denoted by the Greek
letters $\rho$, $\sigma$, $\omega$, ... All others linear operators
(in particular, unbounded operators) will be denoted by the Latin
letters $A$, $B$, $F$, $H$, ...

Denote by $I_{\mathcal{H}}$ the unit operator in a Hilbert space
$\mathcal{H}$ and by $\id_{\mathcal{\H}}$ the identity
transformation of the Banach space $\mathfrak{T}(\mathcal{H})$.

If quantum systems $A$ and $B$ are described by Hilbert spaces  $\H_A$ and $\H_B$ then the bipartite system $AB$ is described by the tensor product of these spaces, i.e. $\H_{AB}\doteq\H_A\otimes\H_B$. A state in $\S(\H_{AB})$ is denoted $\omega_{AB}$, its marginal states $\Tr_{\H_B}\omega_{AB}$ and $\Tr_{\H_A}\omega_{AB}$ are denoted respectively $\omega_{A}$ and $\omega_{B}$.\smallskip

We will use the compactness criterion for subsets of $\S(\H_{AB})$ (see \cite[the Appendix]{AQC}).\smallskip
\begin{lemma}\label{comp}
\textit{A closed subset $\,\S$ of
$\,\S(\H_{AB})$ is compact if and only if
$$
\S_A\doteq\left\{\shs\omega_A\,|\,\omega_{AB}\in\S\shs\right\}\;\textrm{ and }\;\S_B\doteq\left\{\shs\omega_B\,|\,\omega_{AB}\in\S\shs\right\}
$$
are compact subsets of $\,\S(\H_{A})$ and of $\,\S(\H_{B})$ correspondingly.}
\end{lemma}\smallskip

We will also use the following result of the purification
theory.\smallskip
\begin{lemma}\label{p-lemma}
\textit{Let $\mathcal{H}$ and $\mathcal{K}$ be Hilbert spaces such
that $\,\dim\mathcal{H}=\dim\mathcal{K}$. For an arbitrary pure
state $\omega_{0}$ in
$\,\mathfrak{S}(\mathcal{H}\otimes\mathcal{K})$ and an arbitrary
sequence $\{\rho_{k}\}$ of states in $\,\mathfrak{S}(\mathcal{H})$
converging to the state
$\rho_{0}=\mathrm{Tr}_{\mathcal{K}}\omega_{0}$ there exists a
sequence $\{\omega_{k}\}$ of pure states in
$\,\mathfrak{S}(\mathcal{H}\otimes\mathcal{K})$ converging to the
state $\omega_{0}$ such that\break
$\rho_{k}=\mathrm{Tr}_{\mathcal{K}}\omega_{k}$ for all $\,k$.}
\end{lemma}\smallskip

The assertion of Lemma \ref{p-lemma} can be proved by noting that
the infimum in the definition of the Bures distance (or the supremum
in the definition of the Uhlmann fidelity) between two quantum
states can be taken only over all purifications of one state with
fixed purification of the another state and that the convergence of
a sequence of states in the trace norm distance implies its
convergence in the Bures distance \cite{H-SCI, N&Ch}.\smallskip

A \emph{quantum operation} $\,\Phi$ from a system $A$ to a system
$B$ is a completely positive trace non-increasing linear map
$\mathfrak{T}(\mathcal{H}_A)\rightarrow\mathfrak{T}(\mathcal{H}_B)$,
where $\mathcal{H}_A$ and $\mathcal{H}_B$ are Hilbert spaces
associated with the systems $A$ and $B$. A trace preserving quantum
operation is called \emph{quantum channel} \cite{H-SCI,N&Ch}.

The \emph{von Neumann entropy} $H(\rho)=\mathrm{Tr}\eta(\rho)$ of a
state $\rho\in\mathfrak{S}(\mathcal{H})$, where $\eta(x)=-x\log x$,
has the natural extension to the cone
$\mathfrak{T}_{+}(\mathcal{H})$ (cf.\cite{L-2})\footnote{Here and in
what follows $\log$ denotes the natural logarithm.}
\begin{equation*}%\label{ent-ext}
H(\rho)=[\mathrm{Tr}\rho] H\!\left(\frac{\rho}{\mathrm{Tr}
\rho}\right)=\mathrm{Tr}\eta(\rho)-\eta(\mathrm{Tr}\rho),\quad \rho
\in\mathfrak{T}_{+}(\mathcal{H}).
\end{equation*}

Nonnegativity, concavity and lower semicontinuity of the von Neumann
entropy on the cone $\mathfrak{T}_{+}(\mathcal{H})$ follow from the
corresponding properties of this function on the set
$\mathfrak{S}(\mathcal{H})$ \cite{L-2,W}. By definition
$H(\lambda \rho)=\lambda H(\rho),\, \lambda\geq 0$.

The concavity of the von Neumann entropy is supplemented by the
inequality
\begin{equation}\label{w-k-ineq}
H\!\left(\lambda\rho+(1-\lambda)\sigma\right)\leq \lambda
H(\rho)+(1-\lambda)H(\sigma)+h_2(\lambda),\quad \lambda\in[0,1],
\end{equation}
where $h_2(\lambda)=\eta(\lambda)+\eta(1-\lambda)$, valid for any
states $\rho$ and $\sigma$. \smallskip

The \emph{quantum relative entropy} for two operators $\rho$ and
$\sigma$ in $\mathfrak{T}_{+}(\mathcal{H})$ is defined as follows
(cf.\cite{L-2})
$$
H(\rho\,\|\shs\sigma)=\sum_{i=1}^{+\infty}\langle
i|\,\rho\log\rho-\rho\log\sigma+\sigma-\rho\,|i\rangle,
$$
where $\{|i\rangle\}_{i=1}^{+\infty}$ is the orthonormal basis of
eigenvectors of the operator $\rho$ and it is assumed that
$H(\rho\,\|\sigma)=+\infty$ if $\,\mathrm{supp}\rho$ is not
contained in $\mathrm{supp}\shs\sigma$. This definition implies
$H(\lambda\rho\,\|\shs\lambda\sigma)=\lambda
H(\rho\,\|\shs\sigma),\, \lambda\geq0$.

The \emph{quantum mutual information} of a state $\,\omega_{AB}\,$ of an
infinite-dimensional bipartite quantum system $AB$ is defined as follows (cf.\cite{L-mi})
\begin{equation*}%\label{mi-d-2}
I(A\!:\!B)_{\omega}=H(\omega_{AB}\shs\Vert\shs\omega_{A}\otimes
\omega_{B}).
\end{equation*}
We will use the natural extension of this quantity to positive trace-class operators
\begin{equation}\label{mi-ext}
I(A\!:\!B)_{\omega}\doteq [\Tr\shs\omega] I(A\!:\!B)_{\frac{\omega}{\Tr\shs\omega}},\quad \omega\in\T_{+}(\H_{AB}).
\end{equation}
Basic properties of the relative entropy show that $\omega\mapsto
I(A\!:\!B)_{\omega}$ is a lower semicontinuous function on the cone
$\T_{+}(\H_{AB})$ taking values in $[0,+\infty]$.\smallskip

We will use the identity
\begin{equation}\label{sp-ident}
    I(A\!:\!B)_{\omega}+I(B\!:\!C)_{\omega}=2H(\omega_{B})
\end{equation}
valid for any 1-rank operator $\omega\in\T_+(\H_{ABC})$ (with
possible value $+\infty$ in the both sides). If $H(\omega_{A})$,
$H(\omega_{B})$ and $H(\omega_{C})$ are finite then (\ref{sp-ident})
is easily verified by noting that $H(\omega_{A})=H(\omega_{BC})$,
$H(\omega_{B})=H(\omega_{AC})$ and $H(\omega_{C})=H(\omega_{AB})$.
In general case (\ref{sp-ident}) can be proved by approximation (see the proof of Theorem 1 in \cite[the Appendix]{CMI}).\smallskip

Identity (\ref{sp-ident}) implies the upper bound (cf.\cite{MI-B})
\begin{equation}\label{mi-u-b}
I(A\!:\!B)_{\omega}\leq 2\min\{H(\omega_{A}),H(\omega_{B})\}.
\end{equation}

The \emph{quantum conditional entropy}
\begin{equation*}%\label{c-e-d}
H(A|B)_{\omega}=H(\omega_{AB})-H(\omega_B)
\end{equation*}
can be extended to the convex set
$\{\shs\omega_{AB}\,|\,H(\omega_{A})<+\infty\shs\}$ containing
states with $H(\omega_{AB})=H(\omega_B)=+\infty$ by the formula
\begin{equation}\label{c-e-d+}
H_{\mathrm{e}}(A|B)_{\omega}=H(\omega_{A})-I(A\!:\!B)_{\omega}
\end{equation}
preserving all basic properties of the conditional entropy \cite{Kuz} (for more detailed description of this extension see \cite[Sect.5]{CMI}).
In what follows we will denote it by $H(A|B)_{\omega}$ omitting the subscript $\mathrm{e}$.
\smallskip

\begin{lemma}\label{main-l} \emph{Let $\,V_A$ be an operator in
$\B(\H_A)$ such that $\,\|V_A\|\leq1$ and $\omega_{AB}$ be a state with finite $H(\omega_{A})$. Then
\begin{equation}\label{main-r}
0\leq I(A\!:\!B)_{\omega}-I(A\!:\!B)_{\tilde{\omega}}\leq
2\left[ H(\omega_{A})-H(V_A\omega_{A}V^*_A) \right],
\end{equation}
or, equivalently,
$$
|H(A|B)_{\omega}-H(A|B)_{\tilde{\omega}}|\leq
H(\omega_{A})-H(V_A\omega_{A}V^*_A),
$$
where $\,\tilde{\omega}_{AB}=V_A\otimes I_B\,\omega_{AB}V^*_A\otimes I_B\,$ and  $\,H(A|B)$ is the extended conditional entropy defined by (\ref{c-e-d+}).}
\end{lemma}\smallskip

\emph{Proof.} The left inequality in (\ref{main-r})
follows from the monotonicity of the quantum mutual
information under local operations. \smallskip

Let $\omega_{ABC}$ be a purification of
$\omega_{AB}$. Then identity (\ref{sp-ident}) implies
$$
I(A\!:\!B)_{\omega}+I(A\!:\!C)_{\omega}=2H(\omega_A)
$$
and
$$
I(A\!:\!B)_{\tilde{\omega}}+I(A\!:\!C)_{\tilde{\omega}}=2H(\tilde{\omega}_A)=2H(V_A\omega_{A}V^*_A).
$$
Hence
$$
\left[I(A\!:\!B)_{\omega}-I(A\!:\!B)_{\tilde{\omega}}\right]+\left[I(A\!:\!C)_{\omega}-I(A\!:\!C)_{\tilde{\omega}}\right]
=2\left[H(\omega_{A})-H(V_A\omega_{A}V^*_A)\right].
$$
This implies the right inequality in (\ref{main-r}), since $I(A\!:\!C)_{\omega}\geq
I(A\!:\!C)_{\tilde{\omega}}$ by monotonicity of the quantum mutual
information under local operations.\smallskip

By using (\ref{c-e-d+}) it is easy to show that (\ref{main-r}) is equivalent to the second inequality of the lemma. $\square$\medskip

A state $\omega\in\S(\H_{AB})$ is called \emph{separable} if it belongs to the convex closure of the set of all product states $\rho\otimes\sigma$, where
$\rho\in\S(\H_{A})$ and $\sigma\in\S(\H_{B})$. Any separable state $\omega_{AB}$ can be represented as follows
\begin{equation}\label{sep-st}
    \omega_{AB}=\int_X \rho(x)\otimes\sigma(x)\mu_{\omega}(dx),
\end{equation}
where $X$ is a complete separable metric space, $\mu_{\omega}$ is a
Borel probability measure on $X$, $\rho(x)$ and $\sigma(x)$ are
$\mu_{\omega}$-measurable functions on $X$ taking values in
$\mathrm{ext}\shs\S(\H_A)$ and in $\mathrm{ext}\shs\S(\H_B)$
correspondingly \cite{HSW}. If the measure $\mu_{\omega}$ is purely atomic then (\ref{sep-st}) is converted to the countable decomposition
\begin{equation}\label{sep-st+}
    \omega_{AB}=\sum_i\pi_i\rho_i\otimes\sigma_i,
\end{equation}
where $\{\rho_i\}\subset\mathrm{ext}\shs\S(\H_A)$ and
$\{\sigma_i\}\subset\mathrm{ext}\shs\S(\H_B)$
are collections of pure states and
$\{\pi_i\}$ is a probability distribution.\smallskip
\begin{definition}\label{c-d-s-def}
If a separable state $\omega_{AB}$ has representation (\ref{sep-st+}) then it is called \emph{countably decomposable}.
\end{definition}\smallskip
An essential feature of infinite-dimensional bipartite systems consists in existence of separable states which are not countably decomposable \cite{HSW}. Such states called \emph{countably nondecomposable} play important role in this paper (see Remark \ref{c-n-s} below).

\section{On universal infinite-dimensional extension of entanglement monotones}

A central role in quantitative description of entanglement in composite quantum systems is plaid by the notion of entanglement monotones \cite{4H,P&V,V}. In the case of bipartite system $AB$ an \emph{entanglement monotone} $E$ is a nonnegative function
on the set $\S(\H_{AB})$ possessing the following properties:
\begin{enumerate}[\textrm{EM}1)]
\item $\{\shs E(\omega_{AB})=0\;\}\;\Leftrightarrow\;\{\,\textrm{the state}\;\,\omega\;\,
\textrm{is
separable}\,\}$;
\item monotonicity under selective unilocal operations:
\begin{equation*}
E(\omega_{AB})\geq\sum_{k}\pi_{k}E(\omega_{AB}^{k}),\quad
\pi_{k}=\Tr\shs\Phi_k(\omega_{AB}),\;\;
\omega_{AB}^{k}=\pi_{k}^{-1}\Phi_k(\omega_{AB})
\end{equation*}
for any  state $\omega_{AB}$ and any collection $\{\Phi_k\}$ of
unilocal completely positive maps such that $\sum_k\Phi_k$ is a
channel;
\item convexity: $E(\lambda\shs\rho_{AB}+(1-\lambda)\sigma_{AB})\leq
\lambda E(\rho_{AB})+(1-\lambda)E(\sigma_{AB})$.
\end{enumerate}
The convexity of $E$ guarantees that EM2 is equivalent to monotonicity of $E$ under selective LOCC operations \cite{V}.\smallskip

According to \cite{P&V} an entanglement monotone $E$ is called
\textit{entanglement measure} if at any pure state it coincides with the von Neumann entropy of a marginal state, i.e.
\begin{enumerate}[\textrm{EM}1)]
\item [EM4)] $E(\omega_{AB})=H(\omega_A)=H(\omega_B)\,$ for any pure
state $\omega_{AB}$.
\end{enumerate}

Other desirable properties of entanglement monotones are the following:
\begin{enumerate}[\textrm{EM}1)]
\item [EM5)] additivity for product states:
$E(\omega_{AB}\otimes\omega_{A'B'})=E(\omega_{AB})+E(\omega_{A'B'})$, where $E(\omega_{AB}\otimes\omega_{A'B'})$ corresponds to the decomposition $(AA')(BB')$;
\item [EM6)] subadditivity for product states: $E(\omega_{AB}\otimes\omega_{A'B'})\leq E(\omega_{AB})+E(\omega_{A'B'})$, where $E(\omega_{AB}\otimes\omega_{A'B'})$ corresponds to the decomposition $(AA')(BB')$;
\item [EM7)] strong superadditivity: $E(\omega_{(AA')(BB')})\geq
E(\omega_{AB})+E(\omega_{A'B'})$;
\item [EM8)] monogamy: $E(\omega_{A(BC)})\geq
E(\omega_{AB})+E(\omega_{AC})$ \cite{W&K}.
\end{enumerate}

In the finite-dimensional case it is natural to require continuity of an entanglement monotone $E$ on the  set $\S(\H_{AB})$ of all bipartite states. An important role is plaid by the following  stronger property \cite{4H,P&V}:
\begin{enumerate}[\textrm{EM}1)]
\item [EM9)] asymptotic continuity:
$$
\lim_{n\rightarrow+\infty}\frac{E(\rho^n_{AB})-E(\sigma^n_{AB})}{1+\log\dim\H^n_{AB}}=0
$$
for any sequences of states $\rho^n_{AB},\sigma^n_{AB}\!\in\S(\H^n_{AB})\shs$ such that $\|\rho^n_{AB}-\sigma^n_{AB}\|_1$ tends to zero as $n\rightarrow+\infty$.
\end{enumerate}

In the infinite-dimensional case the global continuity
requirement is too restrictive.\footnote{Nevertheless, in infinite-dimensional bipartite systems there exist globally continuous
entanglement monotones \cite[Example 5]{EM}.} Moreover, the discontinuity of the von
Neumann entropy implies discontinuity of any entanglement monotone
possessing property EM4. In this case it seems reasonable to require that an
entanglement monotone (measure) $E$ must be \emph{closed} or \emph{lower semicontinuous}, which means
that
$$
\liminf_{n\rightarrow+\infty}E(\omega_{AB}^{n})\geq E(\omega_{AB}^{0})
$$
for any sequence $\{\omega_{AB}^{n}\}$  converging to a
state $\omega_{AB}^{0}$ or, equivalently, that the set of states
defined by the inequality $E(\omega_{AB})\leq c$ is closed for any $c\geq0$.
This requirement is motivated by the natural physical observation
that entanglement can not be increased by passage to a limit.

From the physical point of view it is also natural to require that
entanglement monotone (measure) must be continuous on subsets of states produced in
a physical experiment, for example, on the set of states with bounded mean energy. One can also consider the corresponding
version of the asymptotic continuity property (see \cite{ESP} and Corollary \ref{c-b-em-c} below). \medskip

Assume now that $E$ is a given entanglement monotone defined on the set of states of
a bipartite system $AB$ composed of subsystems $A$ and $B$ of arbitrary finite dimensions. One can construct its
extension to the set of states of an infinite-dimensional bipartite system $AB$  as follows
\begin{equation}\label{E-def}
    \widehat{E}(\omega_{AB})=\sup_{P_A,P_B}E(P_A\otimes P_B\,\omega_{AB}P_A\otimes
    P_B),\quad \omega_{AB}\in\S(\H_{AB}),
\end{equation}
where the supremum is over all finite rank projectors $P_A\in\B(\H_A)$ and
$P_B\in\B(\H_B)$ and it is assumed that $E$ is extended to all positive
trace class operators by the natural way
$E(\omega_{AB})=[\Tr\omega_{AB}]\shs E\left(\frac{\omega_{AB}}{\Tr\omega_{AB}}\right)$.
\smallskip

Reasonability of this definition is justified by the following
observations.\smallskip

\begin{property}\label{E-prop} \emph{Let $E$ be a continuous entanglement monotone on the set of states of finite-dimensional bipartite system.}\smallskip

A) \emph{$\widehat{E}$ is an unique lower semicontinuous entanglement monotone on the set $\,\S(\H_{AB})$ such that $\widehat{E}(\omega_{AB})=E(\omega_{AB})$ for any state $\,\omega_{AB}$ with finite rank marginal states  $\,\omega_{A}$ and $\,\omega_{B}$.}\smallskip

B) \emph{$\widehat{E}(\omega_{AB})=\lim_{n\rightarrow\infty}E(P^n_A\otimes P^n_B\,\omega_{AB}P^n_A\otimes
    P^n_B)$ for arbitrary sequences $\{P^n_A\}\subset\B(\H_A)$ and
$\{P^n_B\}\subset\B(\H_B)$ of finite rank projectors strongly
converging to the identity operators $I_A$ and $I_B$.}\smallskip

C) \emph{If $E$ possesses one of the above properties EM4-EM8 then $\widehat{E}$ possesses the same property.}
\end{property}\smallskip

\emph{Proof.} A) The lower semicontinuity of $\widehat{E}$ follows
from its definition and the continuity of the function
$\;\omega_{AB}\mapsto E(P_A\otimes P_B\,\omega_{AB}P_A\otimes P_B)\,$ for
any finite rank projectors $P_A$ and $P_B$.\smallskip

To prove  EM1 for $\widehat{E}$ it suffices to note that
separability of $\omega_{AB}$ is equivalent to separability of all
states of the form $\lambda P_A\otimes P_B\,\omega_{AB}P_A\otimes P_B$, $\lambda\in\mathbb{C}$. Properties EM2 and EM3 for $\widehat{E}$ will be proved later.\smallskip

If $\omega_{AB}$ is a state such that
$\,\mathrm{rank}\shs\omega_{A}<+\infty$ and
$\,\mathrm{rank}\shs\omega_{B}<+\infty$ then $\,E(\omega_{AB})\geq
E(P_A\otimes P_B\,\omega_{AB}P_A\otimes P_B)$ for any finite rank
projectors $P_A$ and $P_B$ by monotonicity of $E$ under local
operations (property EM2). This and (\ref{E-def}) imply
$\widehat{E}(\omega_{AB})=E(\omega_{AB})$.\smallskip

B) Since  $E(P^n_A\otimes P^n_B\,\omega_{AB}P^n_A\otimes P^n_B)=\widehat{E}(P_A^n\otimes P_B^n\,\omega_{AB}P_A^n\otimes P_B^n)$ for all $n$, this follows from (\ref{E-def}) and the lower semicontinuity of $\widehat{E}$.

 \smallskip

Now we can prove EM2 for $\widehat{E}$. Let $\omega^n_{AB}=P^n_A\otimes P^n_B\,\omega_{AB}P^n_A\otimes P^n_B$, where $\{P^n_A\}$ and
$\{P^n_B\}$ are any sequences of finite rank projectors strongly
converging to the identity operators $I_A$ and $I_B$. Then for any finite rank projectors $Q_A$ and $Q_B$  the assumed validity of EM2 for $E$ implies
$$
E(\omega^n_{AB})\geq \sum_k E(Q_A\otimes
Q_B\,\Phi_k(\omega^n_{AB})\shs Q_A\otimes Q_B)
$$
for any given collection $\{\Phi_k\}$ of
unilocal completely positive maps such that $\sum_k\Phi_k$ is a
channel. It follows from the continuity of $E$ and assertion B that
$$
\widehat{E}(\omega_{AB})\geq \sum_k E(Q_A\otimes
Q_B\,\Phi_k(\omega_{AB})\shs Q_A\otimes Q_B).
$$
Since this holds for any projectors $Q_A$ and $Q_B$, we have
$$
\widehat{E}(\omega_{AB})\geq \sum_k
\widehat{E}(\Phi_k(\omega_{AB}))=\sum_k
[\Tr\Phi_k(\omega_{AB})]\widehat{E}\left(\frac{\Phi_k(\omega_{AB})}{\Tr\Phi_k(\omega_{AB})}\right)
$$

Assertion B also makes possible to derive convexity of $\widehat{E}$ (property EM3) from the convexity of $E$ and to prove assertion C (since the sequences $\{P^n_A\}$ and
$\{P^n_B\}$ in B can be chosen arbitrarily). $\square$ \medskip

By Proposition \ref{E-prop} the function $\widehat{E}$ is an unique lower semicontinuous extension to the set $\S(\H_{AB})$ of the function $E$ defined on the dense subset
\begin{equation*}%\label{s-f-def}
\S_\mathrm{f}(\H_{AB})\doteq\left\{\shs\omega_{AB}\shs |\shs\max\{\mathrm{rank}\omega_{A},\mathrm{rank}\omega_{B}\}<+\infty\shs\right\}
\end{equation*}
of $\S(\H_{AB})$. By the proof of Proposition \ref{E-prop} the existence and uniqueness of this extension follow from
continuity of the function $E$ and its monotonicity under local operations.\smallskip

Since the above construction can be applied to arbitrary entanglement monotone $E$, we will call the function $\widehat{E}$  an \emph{universal extension} of $E$.\medskip

\textbf{Example: the entanglement of formation.} In the case of finite-dimensional bipartite system $AB$ the entanglement of
formation is defined as the convex roof extension to the set $\S(\H_{AB})$ of the function $\omega_{AB}\mapsto H(\omega_{A})$ on the set $\mathrm{ext}\shs\S(\H_{AB})$ of pure states, i.e.
\begin{equation}\label{ef-def}
  E_{F}(\omega_{AB})=\inf_{\sum_i\pi_i\omega_{AB}^i=\omega_{AB}}\sum_i \pi_iH(\omega^i_{A}),
\end{equation}
where the infimum is over all ensembles $\{\pi_i, \omega_{AB}^i\}$ of pure states with the average state $\omega_{AB}$ \cite{Bennett}.\smallskip

It is well known that $E_F$ is a continuous entanglement measure on $\S(\H_{AB})$ possessing properties EM1-EM4,EM6 and EM9 (and that EM5, EM7 and EM8  do not hold for $E_F$) \cite{H-SCI,W&K,Nielsen}.

In infinite dimensions there are two versions $E_F^{d}$ and $E_F^{c}$ of the entanglement of formation defined, respectively, by using discrete and continuous convex roof extensions, i.e.
$$
E_F^{d}(\omega_{AB})=\!\inf_{\sum_i\!\pi_i\omega^i_{AB}=\omega_{AB}}\sum_i\pi_iH(\omega^i_A),\;\;\;\;\; E_F^{c}(\omega_{AB})=\!\inf_{b(\mu)=\omega_{AB}}\int\! H(\omega_A)\mu(d\omega),
$$
where the first infimum is over all countable convex decompositions of the state $\omega_{AB}$ into pure states and the second one is over all Borel probability measures on the set $\mathrm{ext}\shs\S(\H_{AB})$ with the barycenter $\omega_{AB}$ \cite[Sect.5]{EM}. The discrete version seems more preferable but the assumption $E_F^{d}\neq E_F^{c}$ leads to several problems with this version, in particular, the existence of countably nondecomposable separable states prevents to prove the implication $"\Leftarrow"$ in EM1 for $E_F^{d}$, see the end of Section 4.3.\footnote{In general, the discrete convex roof construction applied to an entropy type function (in the role of $H$) may give a function for which this implication is not valid \cite[Rem.6]{EM}.}

The coincidence of $E_F^{d}$ and $E_F^{c}$ on $\S(\H_{AB})$ is an open question. In \cite{EM} it is shown that $E_F^{d}(\omega_{AB})=E_F^{c}(\omega_{AB})$ for any state $\omega_{AB}$ such that $\min\{H(\omega_{A}),H(\omega_{B}),H(\omega_{AB})\}<+\infty$  and that the function $E_F^{c}$ is lower semicontinuous on $\S(\H_{AB})$. So, Proposition \ref{E-prop} implies $\widehat{E}_F=E_F^{c}$.\smallskip
\begin{corollary}\label{E-prop-c}
\emph{The universal extension $\widehat{E}_F$ of the entanglement of formation $E_F$ (defined by formula (\ref{ef-def})) coincides with the function $E_F^{c}$.}
\end{corollary}
\smallskip

Proposition \ref{E-prop} and Corollary \ref{E-prop-c} show that
$$
\left\{ E_F^{d}=E_F^{c}\right\}\quad\Leftrightarrow\quad \left\{E_F^{d}\;\, \textrm{is lower semicontinuous on} \,\;\S(\H_{AB})\right\}
$$
and provide an alternative proof of properties EM1-EM4 and EM6 for the function
$E_F^{c}$. In fact, $E_F^{c}$ possesses the generalized (continuous) versions of EM2 and EM3 \cite{EM}.

\section{The squashed entanglement}

\subsection{Finite-dimensional case}

The squashed entanglement of a state $\omega_{AB}$ of a finite
dimensional bipartite system $AB$ is defined as follows
\begin{equation}\label{se-def}
  E_{sq}(\omega_{AB})=\textstyle\frac{1}{2}\displaystyle\inf_{\omega_{ABE}}I(A\!:\!B|E),
\end{equation}
where the infimum is over all extensions $\omega_{ABE}$ of the state
$\omega_{AB}$ and
\begin{equation}\label{cmi-def}
I(A\!:\!B|E)_{\omega}=H(\omega_{AE})+H(\omega_{BE})-H(\omega_{E})-H(\omega_{ABE})
\end{equation}
is the conditional mutual information of the state
$\omega_{ABE}$ \cite{C&W,Tucci}. It is essential that the dimension of the
system $E$ in (\ref{se-def}) is assumed to be finite but not bounded
(despite fixed finite dimensions of the systems $A$ and
$B$).\smallskip

The squashed entanglement is the only known entanglement measure possessing all properties EM1-EM9 stated in Section 3.
All these properties excluding EM8,EM9 and the implication $"\Rightarrow"$ in
EM1 are proved in  \cite{C&W}. This implication (called
the faithfulness of $E_{sq}$) is proved in \cite{SE-F}. The monogamy relation EM8 is proved in \cite{W&K}.

Some difficulty concerns the proof of continuity of the squashed
entanglement on $\S(\H_{AB})$. This difficulty is related to unbounded dimension of the system $E$
in definition (\ref{se-def}). The continuity of squashed
entanglement was proved in \cite{C&W} under the conjecture of
validity of the Fannes type continuity bound for quantum conditional
entropy $H(A|B)$ not depending on the dimension of $B$ which was proved later in \cite{A&F}. This
continuity bound also implies the asymptotic continuity of the squashed entanglement.
\medskip

It is also shown in \cite{C&W} that $\,E_{D}(\omega_{AB})\leq
E_{sq}(\omega_{AB})\leq E_{C}(\omega_{AB})\,$ for any state
$\omega_{AB}$, where $E_{D}$ is the distillable entanglement and
$E_{C}$ is the entanglement cost.

The squashed entanglement has an operational interpretation in terms of the protocol of quantum state redistribution \cite{D&J}.
Its interpretation as a distance to highly extendible states is given recently in \cite{L&W-2}.

\subsection{Direct definition of the squashed entanglement in infinite dimensions and its properties}

If $A$ and $B$ are infinite-dimensional systems then we may define
the squashed entanglement by the same formula (\ref{se-def}) in
which \emph{it is necessary} to consider that $E$ is an
infinite-dimensional system as well (see the remark after Lemma \ref{sq-l}
below). The only problem consists in definition of $I(A\!:\!B|E)$,
since formula (\ref{cmi-def}) may contain the uncertainty
$"\infty-\infty"$ even for a state $\omega_{AB}$ with finite marginal
entropies. This problem can be solved by using the extension of
conditional mutual information defined
by one of the equivalent expressions
\begin{equation}\label{cmi-e+}
\!I(A\!:\!B|E)_{\omega}=\sup_{P_A}\left[\shs I(A\!:\!BE)_{Q_A\omega
Q_A}-I(A\!:\!E)_{Q_A\omega Q_A}\shs\right],\; Q_A=P_A\otimes I_{BE},\!
\end{equation}
\begin{equation}\label{cmi-e++}
\!I(A\!:\!B|E)_{\omega}=\sup_{P_B}\left[\shs I(AE\!:\!B)_{Q_B\omega
Q_B}-I(E\!:\!B)_{Q_B\omega Q_B}\shs\right],\; Q_B=P_B\otimes I_{AE},\!
\end{equation}
where the suprema are over all finite rank projectors
$P_A\in\B(\H_A)$ and\break $P_B\in\B(\H_B)$ correspondingly \cite{CMI}.\smallskip

It is shown in \cite[Th.2]{CMI} that expressions (\ref{cmi-e+}) and
(\ref{cmi-e++}) define a lower semicontinuous function on the set
$\S(\H_{ABE})$ possessing all basic properties of conditional mutual
information valid in finite dimensions (including the characterization of a state $\omega_{ABE}$ such that $\,I(A\!:\!B|E)_{\omega}=0\,$ as a short Markov chain in terms of \cite{H&Co}, i.e. as a state such that $\,\omega_{ABE}=\id_A\otimes\Phi(\omega_{AE})$ for some channel $\Phi:E\rightarrow BE$). If one of the marginal
entropies $H(\omega_A)$, $H(\omega_B)$ and $H(\omega_E)$ is finite
then the above extension is given respectively by the explicit formula
\begin{equation}\label{cmi-d+}
I(A\!:\!B|E)_{\omega}=I(A\!:\!BE)_{\omega}-I(A\!:\!E)_{\omega},
\end{equation}
\begin{equation*}%\label{cmi-d++}
I(A\!:\!B|E)_{\omega}=I(AE\!:\!B)_{\omega}-I(E\!:\!B)_{\omega},
\end{equation*}
and
\begin{equation}\label{cmi-d+++}
I(A\!:\!B|E)_{\omega}=I(A\!:\!B)_{\omega}-I(A\!:\!E)_{\omega}-I(B\!:\!E)_{\omega}+I(AB\!:\!E)_{\omega}.
\end{equation}

We will consider in this subsection that
\begin{equation}\label{se-def+}
  E_{sq}(\omega_{AB})=\textstyle\frac{1}{2}\displaystyle\inf_{\omega_{ABE}}I(A\!:\!B|E),\quad \dim\H_E=+\infty
\end{equation}
where $I(A\!:\!B|E)_{\omega}$ is the extended conditional mutual
information described before. Introduce the monotone sequence of functions
\begin{equation}\label{se-def-n}
  E^n_{sq}(\omega_{AB})=\textstyle\frac{1}{2}\displaystyle\inf_{\omega_{ABE}}I(A\!:\!B|E),\quad \dim\H_E\leq n
\end{equation}
pointwise converging to the function
\begin{equation}\label{se-def-s}
  E^*_{sq}(\omega_{AB})=\textstyle\frac{1}{2}\displaystyle\inf_{\omega_{ABE}}I(A\!:\!B|E),\quad \dim\H_E<+\infty.
\end{equation}

In finite-dimensions $E^*_{sq}=E_{sq}$, but the following lemma shows that these functions do not coincide in general.\smallskip

\begin{lemma}\label{sq-l} A) \emph{If $\,I(A\!:\!B)_{\omega}<+\infty$ then $\,E^*_{sq}(\omega_{AB})=E_{sq}(\omega_{AB})<+\infty$.}\smallskip

B) \emph{If $\,I(A\!:\!B)_{\omega}=+\infty$ then $\,E^*_{sq}(\omega_{AB})=+\infty$.}
\end{lemma}
\smallskip

So, if $\omega_{AB}$ is a countably decomposable separable state such that\break
$I(A\!:\!B)_{\omega}=+\infty$, for example, the state
$\omega_{AB}=\sum_{k=1}^{+\infty}\pi_k|k\rangle\langle
k|\otimes|k\rangle\langle k|$, where $\{|k\rangle\}$ is an
orthonormal basis in $\H_A\cong\H_B$ and $\{\pi_k\}$ is a
probability distribution with infinite Shannon entropy, then $\,E^*_{sq}(\omega_{AB})=+\infty$ while $\,E_{sq}(\omega_{AB})=0$ (this follows from Proposition \ref{se-pr}A below). Lemma \ref{sq-l} also shows that $E^*_{sq}=E_{sq}$ if and only if one of the systems $A$ and $B$ is finite-dimensional.\smallskip

\emph{Proof.}  A) Note first that the assumed finiteness of $\,I(A\!:\!B)_{\omega}$ implies finiteness of $E^*_{sq}(\omega_{AB})$ and
of $E_{sq}(\omega_{AB})$.

We will use the inequality
\begin{equation}\label{mi-conc}
\left|I(A\!:\!B|E)_{\lambda\rho+(1-\lambda)\sigma}-\lambda I(A\!:\!B|E)_{\rho}-(1-\lambda)I(A\!:\!B|E)_{\sigma}\right|\leq 2h_2(\lambda)
\end{equation}
valid for any states $\rho_{ABE}$ and $\sigma_{ABE}$ such that $I(A\!:\!B|E)_{\rho}$ and $I(A\!:\!B|E)_{\sigma}$ are finite and any $\lambda\in(0,1)$. If
$\rho_{ABE}$ and $\sigma_{ABE}$ are states with finite marginal entropies then inequality (\ref{mi-conc}) directly follows from
formula (\ref{cmi-def}), concavity of the von Neumann entropy and inequality (\ref{w-k-ineq}). The validity of this inequality for arbitrary
states with finite values of $I(A\!:\!B|E)$ can be proved by using the approximating property for $I(A\!:\!B|E)$ from Theorem 2 in \cite{CMI}.

Let $\varepsilon>0$ be arbitrary, $\omega_{ABE}$ be an extension of $\omega_{AB}$
such that
\begin{equation}\label{I-ineq-1}
I(A\!:\!B|E)_{\omega}<E_{sq}(\omega_{AB})+\varepsilon
\end{equation}
and $\omega_{ABED}$ be a purification of
$\omega_{ABE}$.  Let $\{P^n_D\}\subset\B(\H_D)$ be a sequence of projectors strongly
converging to the operator $I_D$ such that $\mathrm{rank}\shs P^n_D\leq n$ and
$$
\omega^n_{ABED}=(1-\lambda_n)^{-1}I_{ABE}\otimes P^n_D\,\omega_{ABED}\shs I_{ABE}\otimes P^n_D,\quad \lambda_n=1-\Tr P^n_D\omega_{D},
$$
for all $n$. Since $(1-\lambda_n)\omega^n_{ABE}\leq\omega_{ABE}$, we have $\omega_{ABE}=(1-\lambda_n)\omega^n_{ABE}+\lambda_n\tilde{\omega}^n_{ABE}$,
where $\tilde{\omega}^n_{ABE}=\lambda_n^{-1}\left(\omega_{ABE}-(1-\lambda_n)\omega^n_{ABE}\right)$, and hence (\ref{mi-conc}) implies
\begin{equation}\label{I-ineq-2}
  I(A\!:\!B|E)_{\omega}\geq (1-\lambda_n)I(A\!:\!B|E)_{\omega^n}+\lambda_nI(A\!:\!B|E)_{\tilde{\omega}^n}-2h_2(\lambda_n).
\end{equation}
For each $n$ the state $\hat{\omega}^n_{ABD}=(1-\lambda_n)\omega^n_{ABD}+\lambda_n\tilde{\omega}^n_{AB}\otimes\tau_D^n$, where $\tau_D^n$ is a pure state in $\S(P_D^n(\H_D))$, is an extension of $\omega_{AB}$. By using (\ref{mi-conc}), (\ref{I-ineq-2}), the duality relation $I(A\!:\!B|E)_{\omega^n}\!=\!I(A\!:\!B|D)_{\omega^n}$ (cf.\cite{D&J}) and nonnegativity of $I(A\!:\!B|E)_{\tilde{\omega}^n}$ we obtain
$$
\begin{array}{rl}
 I(A\!:\!B|D)_{\hat{\omega}^n}\!\!&\leq\, (1-\lambda_n)I(A\!:\!B|D)_{\omega^n}+\lambda_nI(A\!:\!B|D)_{\tilde{\omega}^n_{AB}\otimes\tau_D^n}+2h_2(\lambda_n)\\\\
&=\,(1-\lambda_n)I(A\!:\!B|E)_{\omega^n}+\lambda_nI(A\!:\!B)_{\tilde{\omega}^n}+2h_2(\lambda_n)\\\\&\leq\, I(A\!:\!B|E)_{\omega}+\lambda_nI(A\!:\!B)_{\tilde{\omega}^n}+4h_2(\lambda_n).
\end{array}
$$
Since $\mathrm{rank}\shs\hat{\omega}^n_{D}\leq n$, this inequality and (\ref{I-ineq-1}) imply
$$
E^n_{sq}(\omega_{AB})\leq E_{sq}(\omega_{AB})+\varepsilon+\lambda_nI(A\!:\!B)_{\tilde{\omega}^n}+4h_2(\lambda_n).
$$
So, to prove that $\,\lim_{n\rightarrow+\infty}E^n_{sq}(\omega_{AB})=E_{sq}(\omega_{AB})\,$ it suffices to show that
\begin{equation}\label{I-lr}
\lim_{n\rightarrow+\infty}\lambda_nI(A\!:\!B)_{\tilde{\omega}^n}=0.
\end{equation}
Since $\omega_{AB}=(1-\lambda_n)\omega^n_{AB}+\lambda_n\tilde{\omega}^n_{AB}$, it follows from (\ref{mi-conc}) that
\begin{equation}\label{I-ineq-3}
I(A\!:\!B)_{\omega}\geq (1-\lambda_n)I(A\!:\!B)_{\omega^n}+\lambda_nI(A\!:\!B)_{\tilde{\omega}^n}-2h_2(\lambda_n).
\end{equation}
Hence, by nonnegativity and lower semicontinuity of the quantum mutual information we have
\begin{equation*}
\lim_{n\rightarrow+\infty}(1-\lambda_n)I(A\!:\!B)_{\omega^n}=I(A\!:\!B)_{\omega}.
\end{equation*}
Since $I(A\!:\!B)_{\omega}<+\infty$, this relation and (\ref{I-ineq-3}) imply (\ref{I-lr}). \smallskip

B) If
$\omega_{ABE}$ is any extension of the state $\omega_{AB}$ such that
$\mathrm{rank}\omega_E<+\infty$ then formula (\ref{cmi-d+++}) and upper bound
(\ref{mi-u-b}) imply $I(A\!:\!B|E)_{\omega}=+\infty$. Hence
$E^*_{sq}(\omega_{AB})=+\infty$. $\square$\smallskip

Consider now properties of the squashed entanglement $E_{sq}$  defined
by formula (\ref{se-def+}). \smallskip

\begin{property}\label{se-pr} \emph{Let $\,\S_\mathrm{*}\doteq\left\{\shs\omega_{AB}\,|\,\min\{H(\omega_{A}),H(\omega_{B}),H(\omega_{AB})\}<+\infty\shs\right\}$.}
\begin{enumerate}[A)]
    \item \emph{If $\,E_{sq}(\omega_{AB})=0\,$ then $\,\omega_{AB}$ is a separable state, the converse implication holds if $\,\omega_{AB}$ is a state in $\,\mathrm{conv}(\S_\mathrm{*}\cup\shs\S_\mathrm{cd})$, where $\S_\mathrm{cd}$ is the set of countably decomposable separable states (Def.\ref{c-d-s-def}).}\footnote{$\,\mathrm{conv}(\S_\mathrm{*}\cup\shs\S_\mathrm{cd})$ is the convex hull of $\S_\mathrm{*}\cup\shs\S_\mathrm{cd}$. Obstacles preventing to prove that  $\,E_{sq}(\omega_{AB})=0\,$ for countably nondecomposable separable states $\,\omega_{AB}\,$ with infinite marginal entropies are considered in Remark \ref{c-n-s} below.}
        \item \emph{The function $E_{sq}$ possesses the above properties
EM2-EM8.}
    \item \emph{The function $E_{sq}$ is lower-semicontinuous
    on the set $\,\S_\mathrm{*}$ and coincides on this set with the function $E_{sq}^*$ defined by formula (\ref{se-def-s}).}
    \item \emph{The function $E_{sq}$ is continuous on any subset of
$\,\S(\H_{AB})$ on which \textbf{one} of the functions $\,\omega_{AB}\mapsto H(\omega_{A})$ and $\,\omega_{AB}\mapsto H(\omega_{B})$ is continuous}.
\end{enumerate}
\end{property}

Assertion D in Proposition \ref{se-pr} is essentially strengthened in Section 5.\smallskip

\emph{Proof.} B) Properties EM2-EM8 are proved by the same arguments as in the
finite-dimensional case (see \cite{C&W,W&K}) with obvious modifications using properties of the extended conditional mutual information stated in \cite[Th.2]{CMI}.\smallskip

C) Take increasing sequences $\{P^n_A\}\subset\B(\H_A)$ and
$\{P^n_B\}\subset\B(\H_B)$ of finite rank projectors strongly
converging to the identity operators $I_A$ and $I_B$. For each $n$
consider the functions
$$
f_n(\omega_{AB})=\textstyle\frac{1}{2}\displaystyle\inf_{\omega_{ABE}}\shs
I(A\!:\!B|E)_{Q^n_A\omega Q^n_A}\quad\textrm{and}\quad \displaystyle
g_n(\omega_{AB})=\textstyle\frac{1}{2}\displaystyle\inf_{\omega_{ABE}}\shs
I(A\!:\!B|E)_{Q^n_B\omega Q^n_B},
$$
where $Q^n_A=P^n_A\otimes I_{BE}$, $Q^n_B=P^n_B\otimes I_{AE}$ and the infima are over all extensions $\shs\omega_{ABE}$ of the state
$\shs\omega_{AB}$.

We will show first that these functions are continuous on
$\S(\H_{AB})$ for all $n$. By symmetry it suffices to prove the
continuity of $f_n$. Let $\omega^1_{AB}$ and
$\omega^2_{AB}$ be states such that
$\|\omega^2_{AB}-\omega^1_{AB}\|_1\leq\varepsilon\leq1$. By repeating
the arguments from the proof of continuity of $E_{sq}$ in \cite{C&W} we obtain
\begin{equation}\label{sq-rep}
f_n(\omega^k_{AB})=\textstyle\frac{1}{2}\displaystyle\inf_{\Lambda}I(A\!:\!B|E)_{Q^n_A\id_{AB}\otimes\Lambda(\omega^k_{ABC})
Q^n_A},\quad k=1,2,
\end{equation}
where $\omega^1_{ABC}$  and $\omega^2_{ABC}$ are purifications of
$\omega^1_{AB}$ and of $\omega^2_{AB}$ such that
$$\|\omega^2_{ABC}-\omega^1_{ABC}\|_1\leq2\sqrt{\varepsilon},$$
and the infimum is over all quantum channels
$\Lambda:\T(\H_C)\rightarrow\T(\H_E)$.\smallskip

For given $\Lambda$ let
$\omega^1_{ABE}=\id_{AB}\otimes\Lambda(\omega^1_{ABC})$ and
$\omega^2_{ABE}=\id_{AB}\otimes\Lambda(\omega^2_{ABC})$. Then
$\|Q^n_A \omega^2_{ABE}Q^n_A-Q^n_A\omega^1_{ABE}Q^n_A\|_1\leq2\sqrt{\varepsilon}$.
Let $\lambda_k=\Tr Q^n_A \omega^k_{ABE}$, $\,\lambda_k=1,2$.
Note that
\begin{equation}\label{l-d}
  |\lambda_2-\lambda_1|\leq\|\omega^2_{AB}-\omega^1_{AB}\|_1\leq\varepsilon
\end{equation}
and that
\begin{equation}\label{cmi-u-b}
I(A\!:\!B|E)_{Q^n_A\omega^k Q^n_A}
\leq2
\lambda_k\log\mathrm{rank}P^n_A,\quad k=1,2.
\end{equation}
Inequality (\ref{cmi-u-b}) follows from representation (\ref{cmi-d+}) and the upper bound (\ref{mi-u-b}). \smallskip

If $\lambda_1\lambda_2=0$ then (\ref{l-d}) and (\ref{cmi-u-b}) imply
\begin{equation}\label{rt-1}
\Delta_{12}\doteq\left|I(A\!:\!B|E)_{Q^n_A\omega^2 Q^n_A}-I(A\!:\!B|E)_{Q^n_A\omega^1
Q^n_A}\right|\leq2\varepsilon
\log\mathrm{rank}P^n_A.
\end{equation}

If $\lambda_1\lambda_2\neq0$ denote the states $\lambda_1^{-1}Q^n_A \omega^1_{ABE}Q^n$ and $\lambda_2^{-1}Q^n_A \omega^2_{ABE}Q^n$ respectively by $\hat{\omega}^2_{ABE}$ and $\hat{\omega}^1_{ABE}$. It follows from (\ref{l-d}) and (\ref{cmi-u-b}) that
\begin{equation}\label{rt-2}
\begin{array}{rl}
\Delta_{12}\!&=\;|\lambda_2 I(A\!:\!B|E)_{\hat{\omega}^2}-\lambda_1 I(A\!:\!B|E)_{\hat{\omega}^1}|\\\\
&\leq\;\lambda_2|I(A\!:\!B|E)_{\hat{\omega}^2}-I(A\!:\!B|E)_{\hat{\omega}^1}|+2\varepsilon
\log\mathrm{rank}P^n_A.
\end{array}
\end{equation}
Since
$$
\lambda_2\|\hat{\omega}^2_{ABE}-\hat{\omega}^1_{ABE}\|_1=\|\lambda_2\hat{\omega}^2_{ABE}-\lambda_1\hat{\omega}^1_{ABE}\|_1+|\lambda_2-\lambda_1|
\leq\tilde{\varepsilon}\doteq2\sqrt{\varepsilon}+\varepsilon,
$$
the Fannes type continuity bound for the conditional mutual information (Corollary 8 in \cite{CMI}) implies
\begin{equation}\label{rt-3}
\lambda_2|I(A\!:\!B|E)_{\hat{\omega}^2}-I(A\!:\!B|E)_{\hat{\omega}^1}|\leq \tilde{\varepsilon}
\log\mathrm{rank}P^n_A+4\lambda_2\theta\left(\frac{\tilde{\varepsilon}}{2\lambda_2}\right),
\end{equation}
where $\theta(x)=(1+x)h_2\!\left(\frac{x}{1+x}\right)$. \smallskip

It is easy to see that $\lambda_2\theta\left(\frac{\tilde{\varepsilon}}{2\lambda_2}\right)$ tends to zero as $\,\tilde{\varepsilon}\rightarrow0$
uniformly on $\lambda_2\in(0,1]$. Hence, (\ref{rt-1})-(\ref{rt-3}) show that
$\Delta_{12}\leq \tau(\varepsilon)$  for arbitrary $\lambda_1,\lambda_2$, where $\tau(\varepsilon)$ is a function
tending to zero as $\,\varepsilon\rightarrow0$. Since this holds for any quantum channel $\Lambda$, we conclude from
(\ref{sq-rep}) that
$|f_n(\omega^2_{AB})-f_n(\omega^1_{AB})|\leq
\frac{1}{2}\tau(\varepsilon)$. Thus, the function $f_n$ is
continuous.\smallskip

The continuity of the functions $f_n$ and
$g_n$ implies  continuity of the function
$$
h_{n}(\omega_{AB})\doteq\max\left\{f_n(\omega_{AB}),g_n(\omega_{AB})\right\}.
$$
So, to prove the lower semicontinuity of $E_{sq}$ on $\S_*$ it suffices to show
that
\begin{equation}\label{sup-c}
 \sup_n h_{n}(\omega_{AB})=E_{sq}(\omega_{AB})
\end{equation}
for any
state $\omega_{AB}$ in $\S_*$. If $\omega_{AB}$ is a state such that either $H(\omega_{A})<+\infty$ or
$H(\omega_{B})<+\infty$ then (\ref{sup-c}) follows from Lemma \ref{main-l+} below,
since it implies
\begin{equation}\label{tmp-e}
E_{sq}(\omega_{AB})-f_n(\omega_{AB})\leq
\left[H(\omega_{A})-H(P^n_A\omega_{A}P^n_A)\right]
\end{equation}
if $H(\omega_{A})<+\infty$ and
$$
E_{sq}(\omega_{AB})-g_n(\omega_{AB})\leq
\left[H(\omega_{B})-H(P^n_B\,\omega_{B}P^n_B)\right]
$$
if $H(\omega_{B})<+\infty$. Since
$\lim_{n\rightarrow\infty}H(P^n_X\,\omega_{X}P^n_X)=H(\omega_{X})$,
$X=A,B$, by Simon's convergence theorem \cite[the Appendix]{Simon} and
$$
E_{sq}(\omega_{AB})-h_{n}(\omega_{AB})=\min\{E_{sq}(\omega_{AB})-f_n(\omega_{AB}),E_{sq}(\omega_{AB})-g_n(\omega_{AB})\}
$$
by definition of $h_{n}$, we have $\,\lim_{n\rightarrow\infty}h_{n}(\omega_{AB})=E_{sq}(\omega_{AB})$.\smallskip

If $\omega_{AB}$ is a state such that $H(\omega_{A})=H(\omega_{B})=+\infty$ but
$H(\omega_{AB})<+\infty$ then $\,I(A\!:\!B)_{\omega}=+\infty$. Since $f_n(\omega_{AB})\leq E_{sq}(\omega_{AB})$ for all $n$, to prove (\ref{sup-c}) it suffices to show that
\begin{equation}\label{f-lim-rel}
\lim_{n\rightarrow\infty}f_n(\omega_{AB})=+\infty.
\end{equation}
The lower semicontinuity of $\,I(A\!:\!B)$ and Simon's convergence theorem imply respectively
$$
\lim_{n\rightarrow\infty} I(A\!:\!B)_{Q^n_A\omega Q^n_A}=I(A\!:\!B)_{\omega}=+\infty
$$
and
$$
\lim_{n\rightarrow\infty} H(Q^n_A\omega_{AB} Q^n_A)=H(\omega_{AB})<+\infty.
$$
So, limit relation (\ref{f-lim-rel}) follows from the inequality
\begin{equation}\label{u-ineq}
  I(A\!:\!B|E)_{\omega}\geq I(A\!:\!B)_{\omega}-2H(\omega_{AB}),
\end{equation}
valid for any state $\omega_{ABE}$ with finite $H(\omega_{AB})$. If
$\,I(A\!:\!BE)_{\omega}<+\infty\,$ then (\ref{u-ineq}) is proved by using monotonicity of the quantum mutual information under partial trace (cf.\cite{C&W}): $$
\begin{array}{rl}
I(A\!:\!B|E)_{\omega}\!\!&=\,I(A\!:\!BE)_{\omega}-I(A\!:\!E)_{\omega}\\\\&\geq\,I(A\!:\!B)_{\omega}-I(AB\!:\!ED)_{\omega}
\geq I(A\!:\!B)_{\omega}-2H(\omega_{AB}),
\end{array}
$$
where it is assumed that $\omega_{ABED}$ is a purification of $\omega_{ABE}$. The validity of inequality (\ref{u-ineq}) for arbitrary
states $\omega_{ABE}$ with finite $H(\omega_{AB})$ can be proved by using the approximating property for $I(A\!:\!B|E)$ from Theorem 2 in \cite{CMI}.\smallskip

The coincidence of $E_{sq}$ and $E^*_{sq}$ on $\S_*$ follows from Lemma \ref{sq-l} and the above observation showing that $E_{sq}(\omega_{AB})=+\infty$ for any state $\omega_{AB}$ such that $H(\omega_{A})=H(\omega_{B})=+\infty$ and
$H(\omega_{AB})<+\infty$.\smallskip

D) Assume the function $\omega_{AB}\mapsto H(\omega_{A})$ is
continuous on a subset $\S_0$ of $\S(\H_{AB})$. By Dini's lemma the
increasing sequence of continuous functions $\omega_{AB}\mapsto
H(P^n_A\,\omega_{A}P^n_A)$  converges to the continuous
function $\omega_{AB}\mapsto H(\omega_{A})$ uniformly on any compact subset of
$\S_0$. So, inequality (\ref{tmp-e}) shows that the sequence of
continuous functions $\,f_n$ uniformly converges to
the function $E_{sq}$ on any compact subset of $\S_0$.
Hence the function $E_{sq}$ is continuous on $\S_0$.

A) Assume $\,E_{sq}(\omega_{AB})=0$. Take any sequences
$\{P^n_A\}\subset\B(\H_A)$ and $\{P^n_B\}\subset\B(\H_B)$ of finite
rank projectors strongly converging to the identity operators $I_A$
and $I_B$. By monotonicity of the conditional mutual information
under local operations we have
$\,E_{sq}(\omega^n_{AB})=0$ for all $n$, where $\omega^n_{AB}=[\Tr P^n_A\otimes
P^n_B\,\omega_{AB}]^{-1}P^n_A\otimes
P^n_B\,\omega_{AB}P^n_A\otimes P^n_B$. By the
faithfulness of the squashed entanglement in finite dimensions
(proved in \cite{SE-F}) all the states $\omega^n_{AB}$ are
separable. So, the state $\omega_{AB}$ is separable (as a limit of a
sequence of separable states).

If $\,\omega_{AB}$ is a separable state in $\S_{\mathrm{cd}}$ having representation (\ref{sep-st+}) then it can be extended to a short
Markov chain as follows (cf.\cite{C&W})
\begin{equation}\label{c-d-ext}
\omega_{ABE}=\sum_{i}\pi_i\shs\rho_i\otimes\sigma_i\otimes|i\rangle \langle
i|,
\end{equation}
where $\{|i\rangle\}$ is an orthonormal basis in some Hilbert space
$\H_E$. Hence, $E_{sq}(\omega_{AB})=0$

If $\,\omega_{AB}$ is a separable state in $\S_{*}$ then it can be represented as a limit of a sequence $\{\omega^n_{AB}\}$  of separable states in $\S_{*}\cap\S_{\mathrm{cd}}$ (for examples, separable states having finite rank marginal states). Since $E_{sq}(\omega^n_{AB})=0$ for all $n$, the lower semicontinuity of $E_{sq}$ on $\S_{*}$ (assertion C) implies $E_{sq}(\omega_{AB})=0$.

If $\,\omega_{AB}$ is a convex mixture of separable states in $\S_{*}$ and in $\S_{\mathrm{cd}}$ then
the convexity of $E_{sq}$ (assertion B) implies $E_{sq}(\omega_{AB})=0$. $\square$ \medskip

\begin{lemma}\label{main-l+} \emph{Let $\,V_A$ be an operator in
$\B(\H_A)$ such that $\,\|V_A\|\leq1$ and $\omega_{ABE}$ be a state with finite $H(\omega_{A})$. Then
\begin{equation}\label{cmi-s-r}
0\leq I(A\!:\!B|E)_{\omega}-I(A\!:\!B|E)_{\tilde{\omega}}\leq
2\left[H(\omega_{A})-H(V_A\omega_{A}V^*_A)\right],
\end{equation}
and hence
$$
-2\delta H(V_A\omega_{A}V^*_A)\leq I(A\!:\!B|E)_{\omega}-I(A\!:\!B|E)_{\frac{\tilde{\omega}}{\Tr\tilde{\omega}}}\leq
2\left[H(\omega_{A})-H(V_A\omega_{A}V^*_A)\right],
$$
where $\,\tilde{\omega}_{ABE}=V_A\otimes I_{BE}\shs\omega_{ABE}V^*_A\otimes
I_{BE}$ and $\,\delta=\frac{1-\Tr\tilde{\omega}}{\Tr\tilde{\omega}}$.}
\end{lemma}\smallskip

\emph{Proof.} The left inequality in (\ref{cmi-s-r}) follows from
the  monotonicity of the conditional mutual information under
local operations. \smallskip

To prove the right inequality in (\ref{cmi-s-r}) it suffices to note that
$$
I(A\!:\!B|E)_{\omega}=H(A|E)_{\omega}-H(A|BE)_{\omega}
$$
for any $\omega\in\T_+(\H_{ABE})$ with finite $H(\omega_A)$, where $H(A|X)$ is the extended conditional entropy defined in (\ref{c-e-d+}), and to apply Lemma \ref{main-l} twice.

The second inequality is easily derived from the first one by noting that
$$
0\leq [\Tr\shs\tilde{\omega}]\shs I(A\!:\!B|E)_{\frac{\tilde{\omega}}{\Tr\tilde{\omega}}}=I(A\!:\!B|E)_{\tilde{\omega}}\leq2H(V_A\omega_{A}V^*_A).\,\square
$$

\begin{remark}\label{c-n-s} We cannot prove that $E_{sq}(\omega_{AB})=0\,$ for any separable state $\omega_{AB}$, since we cannot extend a countably nondecomposable separable state to a Markov chain. It is easy to see that the integral analog of formula (\ref{c-d-ext})  (in which the basis $\{|i\rangle\langle i |\}$ is replaced by a basis $\{|x\rangle\langle x|\}_{x\in X}$ of nonseparable Hilbert space $\H_E$) produces a non-normal state on $\B(\H_{ABE})$.
If $\omega_{AB}$ is a countably nondecomposable separable state such that either $H(\omega_{A})$ or $H(\omega_{B})$ is finite then  we can prove that $\,E_{sq}(\omega_{AB})=0\,$ by approximation but we cannot prove that  the infimum in the definition of $\,E_{sq}(\omega_{AB})\,$ is attained at some state $\omega_{ABE}$.\smallskip

Thus, we have faced with the \textbf{interesting question}: \emph{Can a countably nondecomposable separable state $\omega_{AB}$ be extended to a Markov chain $\omega_{ABE}$?}\smallskip

Note that any such (hypothetical) extension would be a Markov chain  having no
representation described by Hayden, Jozsa, Petz and Winter in \cite{H&Co} characterizing Markov chains in finite-dimensional tripartite systems (it is easy to see that $\omega_{AB}$ is a countably decomposable separable state
for any Markov chain $\omega_{ABE}$ having such representation).
\end{remark}\smallskip

\subsection{The universal extension of squashed entanglement}

According to Section 3 the universal extension of squashed entanglement is defined by the formula
\begin{equation}\label{Esq-def}
    \widehat{E}_{sq}(\omega_{AB})=\sup_{P_A,P_B}E_{sq}(P_A\otimes P_B\,\omega_{AB}P_A\otimes
    P_B),
\end{equation}
where the supremum is over all finite rank projectors $P_A$ and
$P_B$ and $E_{sq}$ is the natural (homogeneous) extension of the finite-dimensional squashed entanglement defined by (\ref{se-def}) to the cone $\T_+(\H_{AB})$.
\smallskip

Properties of the function $\widehat{E}_{sq}$ and relations between this function and the  infinite-dimensional squashed entanglement $E_{sq}$ defined in Section 4.2 are presented in the following proposition.
\smallskip

\begin{property}\label{two-def} A) \emph{$\widehat{E}_{sq}$ is an unique lower semicontinuous entanglement measure on $\S(\H_{AB})$
coinciding
with the "finite-dimensional" squashed entanglement on the set
$\,\S_\mathrm{f}\doteq\{\omega_{AB}\,|\,\max\{\mathrm{rank}\omega_{A},\mathrm{rank}\omega_{B}\}<+\infty\}$;}\smallskip

B) \emph{The function $\,\widehat{E}_{sq}$ possesses properties EM1-EM8;} \smallskip

C) \emph{$\widehat{E}_{sq}(\omega_{AB})\leq E_{sq}(\omega_{AB})$, where $E_{sq}(\omega_{AB})$ is defined by formula (\ref{se-def+}), for any state $\omega_{AB}$
and $\,\widehat{E}_{sq}(\omega_{AB})=E_{sq}(\omega_{AB})\,$ for a state $\omega_{AB}$ in}
\begin{equation}\label{star}
\S_\mathrm{*}\doteq\{\shs\omega_{AB}\,|\,\min\{H(\omega_{A}),H(\omega_{B}),H(\omega_{AB})\}<+\infty\shs\}.
\end{equation}

D) \emph{$\widehat{E}_{sq}=E_{sq}$ if and only if
$\,E_{sq}$ is lower semicontinuous on $\,\S(\H_{AB})$.}
\end{property} \smallskip

\emph{Proof.} Assertions A,B,D  and the inequality $\,\widehat{E}_{sq}(\omega_{AB})\leq E_{sq}(\omega_{AB})$  directly follow from Propositions \ref{E-prop} and \ref{se-pr}.

Let  $\shs\omega_{AB}\in\S_\mathrm{*}$ and
$\,
\omega^n_{AB}=\left[\Tr P^n_A\otimes P^n_B\,\omega_{AB}\right]^{-1} P^n_A\otimes P^n_B\,\omega_{AB}P^n_A\otimes
    P^n_B\in\S_\mathrm{f},
$
where  $\{P^n_A\}$ and $\{P^n_B\}$ are  any sequences of finite rank projectors strongly converging to the operators $I_A$ and $I_B$.
By Proposition \ref{se-pr}  the function
$E_{sq}$ is lower semicontinuous on $\S_\mathrm{*}$ and monotone under
local operations. Hence, $E_{sq}(\omega_{AB})=\lim_{n\rightarrow\infty}E_{sq}(\omega^n_{AB})$.
This and Proposition \ref{E-prop}B  imply
$E_{sq}(\omega_{AB})=\widehat{E}_{sq}(\omega_{AB})$. $\square$.
 \smallskip

Relations between $E_{sq}$ and $\widehat{E}_{sq}$ are very similar to the relations between the infinite-dimensional versions $E_F^{d}$ and $E_F^{c}=\widehat{E}_{F}$ of the entanglement of formation considered at the end of Section 3 (Corollary \ref{E-prop-c}). Indeed (cf.\cite{EM}),
\begin{itemize}
  \item $E_F^{c}$ is a unique lower semicontinuous entanglement measure on $\S(\H_{AB})$ coinciding with the "finite-dimensional" entanglement of formation $E_F$ on the set
$\S_\mathrm{f}$ and inheriting all basic properties of $E_F$;
  \item $E_F^{d}(\omega_{AB})\geq E_F^{c}(\omega_{AB})$ for any state $\omega_{AB}$;
  \item the equality $E_F^{d}(\omega_{AB})=E_F^{c}(\omega_{AB})$ is proved for any state $\omega_{AB}$ in the set $\S_\mathrm{*}$ defined in (\ref{star});
  \item the equality $\,E^d_{F}(\omega_{AB})=0\,$ is proved for any separable state $\,\omega_{AB}$ in $\,\mathrm{conv}(\S_\mathrm{*}\cup\shs\S_\mathrm{cd})$, where $\S_\mathrm{cd}$ is the set of countably decomposable separable states;
  \item $E_F^{d}=E_F^{c}\,$ if and only if $\,E_F^{d}$  is lower semicontinuous on $\S(\H_{AB})$.
\end{itemize}

Similar to the function $E_{sq}$, the equality $\,E_F^{d}(\omega_{AB})=0\,$ \emph{is not proved} (as far as I know) for a countably nondecomposable separable state $\omega_{AB}$ such that $H(\omega_{A})=H(\omega_{B})=+\infty$.\footnote{So, strictly speaking, the both functions $E_{sq}$ and $E_F^{d}$ can not be considered as entanglement measures on $\S(\H_{AB})$.}
\smallskip

In finite dimensions $E_{sq}(\omega_{AB})\leq E_F(\omega_{AB})$ for any state $\omega_{AB}$ \cite[Pr.5]{C&W}. This relation is generalized as follows.
\begin{corollary}\label{sq-ef-ineq} \emph{For any state $\omega_{AB}$ of an infinite-dimensional bipartite system the following relations hold}
$$
E_{sq}(\omega_{AB})\leq E_F^d(\omega_{AB}),\quad \widehat{E}_{sq}(\omega_{AB})\leq E_F^c(\omega_{AB}).
$$
\end{corollary}

\emph{Proof.} The first inequality follows from the representation (cf.\cite{C&W})
\begin{equation}\label{ef-ref}
  E^d_{F}(\omega_{AB})=\textstyle\frac{1}{2}\displaystyle\inf_{\hat{\omega}_{ABE}}I(A\!:\!B|E)_{\hat{\omega}},
\end{equation}
where the infimum is over all extensions of the state $\omega_{AB}$ having the form
\begin{equation*}
  \hat{\omega}_{ABE}=\sum_i \pi_i\omega^i_{AB}\otimes|i\rangle\langle i|,\quad \rank\shs\omega^i_{AB}=1.
\end{equation*}

Since $E_F^c=\widehat{E}_F$ by Corollary \ref{E-prop-c}, the second inequality follows from the definitions
of the functions  $\widehat{E}_{sq}$ and $\widehat{E}_F$. $\square$ \smallskip

\textbf{Note:} we can not assert that $E_{sq}(\omega_{AB})\leq E_F^c(\omega_{AB})$ until it is not proved that either
$E_F^d(\omega_{AB})=E_F^c(\omega_{AB})$ or $E_{sq}(\omega_{AB})=\widehat{E}_{sq}(\omega_{AB})$. \medskip

It is shown in \cite[Sect.5]{EM} that if $\,E_F^{d}\neq E_F^{c}\,$ then the function $E_F^{d}$ demonstrates properties which seems non-adequate for entanglement measure. The same arguments can be repeated for the function $E_{sq}$ if we assume that $E_{sq}(\omega_{AB})>\widehat{E}_{sq}(\omega_{AB})$ for some state $\omega_{AB}$. So, the function $\widehat{E}_{sq}$ seems to be more preferable candidate on the role of infinite-dimensional squashed entanglement (until it is not proved that $E_{sq}=\widehat{E}_{sq}$). \smallskip

From the physical point of view possible noncoincidences of $E_{sq}$ with $\widehat{E}_{sq}$ and of $E^d_{F}$ with $E^c_{F}$ are not too essential, since $\,E_{sq}(\omega_{AB})=\widehat{E}_{sq}(\omega_{AB})\,$ and $\,E^d_{F}(\omega_{AB})=E^c_{F}(\omega_{AB})\,$ for any state $\,\omega_{AB}$
with finite energy provided the Hamiltonian of one of the subsystems A and B satisfies some regularity condition (see Corollary \ref{two-def-c+} in Section 5.2).

\section{On continuity of the squashed entanglement}

\subsection{General continuity condition and its corollaries}

Proposition \ref{se-pr}D states that
\begin{equation}\label{H-c-c}
\lim_{k\rightarrow\infty} H(\omega^k_{X})=H(\omega^0_{X})<+\infty\;
\Rightarrow\;
\lim_{k\rightarrow\infty}E_{sq}(\omega^k_{AB})=E_{sq}(\omega^0_{AB})<+\infty,
\end{equation}
where $X$ is either $A$ or $B$, for a sequence $\{\omega^k_{AB}\}$
converging to a state $\omega^0_{AB}$. In this case $\,E_{sq}(\omega^k_{AB})=\widehat{E}_{sq}(\omega^k_{AB})$ for all $k$ (by Proposition \ref{two-def}).
Exactly the same continuity condition holds for the entanglement of formation \cite[Pr.8]{EM}.\smallskip

In fact, a stronger result relating  continuity of the squashed entanglement with continuity of the quantum mutual information is valid (which seems more natural than condition (\ref{H-c-c}) from the physical point of view).\smallskip

\begin{property}\label{g-c-c} \emph{Let $\{\omega^k_{AB}\}$ be a sequence  converging to a state $\omega^0_{AB}$.}\smallskip

A) \emph{If $\,\min\{H(\omega^0_{A}),H(\omega^0_{B})\}<+\infty$ then the following properties $\mathrm{(i)\textrm{-}(iv)}$ are equivalent and imply $\mathrm{(v)}$:}
\begin{enumerate}[(i)]
  \item \emph{$\displaystyle\lim_{k\rightarrow\infty} I(A\!:\!B)_{\omega^k}=I(A\!:\!B)_{\omega^0}$;}
  \item \emph{$\displaystyle\lim_{k\rightarrow\infty}E^n_{sq}(\omega^k_{AB})=E^n_{sq}(\omega^0_{AB})\,$ for some $\,n$;}\footnote{The function $E_{sq}^n$ is defined in (\ref{se-def-n}).}
  \item \emph{$\displaystyle\lim_{k\rightarrow\infty}E^n_{sq}(\omega^k_{AB})=E^n_{sq}(\omega^0_{AB})\,$ for all $\,n$;}
    \item \emph{$\displaystyle\lim_{k\rightarrow\infty}E_{sq}(\omega^k_{AB})=E_{sq}(\omega^0_{AB})\,$ and $\;\displaystyle\lim_{n\rightarrow\infty}\sup_{k\geq k_*}\!\left[E^n_{sq}(\omega^k_{AB})-E_{sq}(\omega^k_{AB})\right]=0$ for sufficiently large $k_*$;}
  \item \emph{$\displaystyle\lim_{k\rightarrow\infty}\widehat{E}_{sq}(\omega^k_{AB})=\widehat{E}_{sq}(\omega^0_{AB})$.}
\end{enumerate}
\emph{The condition $\,\min\{H(\omega^0_{A}),H(\omega^0_{B})\}<+\infty$ implies $E_{sq}(\omega^0_{AB})=\widehat{E}_{sq}(\omega^0_{AB})$ and finiteness of the above limits (but it does not imply $E_{sq}(\omega^k_{AB})=\widehat{E}_{sq}(\omega^k_{AB})$).}

\medskip
B) \emph{If $\shs H(\omega^0_{A})=H(\omega^0_{B})=+\infty$ and $\shs H(\omega^0_{AB})<+\infty$ then $\mathrm{(i)},\mathrm{(iii)}$, the first part of $\mathrm{(iv)}$ and $\mathrm{(v)}$ hold as infinite limits.}

\medskip
C) \emph{If $\,I(A\!:\!B)_{\omega^0}<+\infty$ and $\,\lambda_k\omega^k_{AB}\leq\omega^0_{AB}$ for all $k$, where $\{\lambda_k\}$ is a sequence of positive numbers converging to $1$,
then  $\mathrm{(i)}\textit{-}\mathrm{(iv)}$ hold.}\smallskip
\end{property}\medskip

\begin{remark}\label{g-c-c-rmk-1} The main assertion of Proposition \ref{g-c-c}A is the implication
\begin{equation}\label{g-c-c-r}
\lim_{k\rightarrow\infty} I(A\!:\!B)_{\omega^k}=I(A\!:\!B)_{\omega^0}\quad\Rightarrow\quad \lim_{k\rightarrow\infty}E(\omega^k_{AB})=E(\omega^0_{AB})
\end{equation}
for $E=E_{sq},\widehat{E}_{sq}$. It strengthens condition (\ref{H-c-c}), since Theorem 1A and Example 1 in \cite{CMI} show that
\begin{equation*}
\lim_{k\rightarrow\infty} H(\omega^k_{X})=H(\omega^0_{X})<+\infty\;\;
\Rightarrow\;\;
\lim_{k\rightarrow\infty}I(A\!:\!B)_{\omega^k_{AB}}=I(A\!:\!B)_{\omega^0_{AB}}<+\infty,
\end{equation*}
for any sequence $\{\omega^k_{AB}\}$ converging to a state $\omega^0_{AB}$, where $X$ is either $A$ or $B$, and that the converse implication is not valid.\smallskip

The main advantage of  continuity condition (\ref{g-c-c-r}) in contrast to (\ref{H-c-c})
consists in the fact that \emph{local continuity of quantum mutual information is preserved by local operations} (in contrast to local continuity of marginal entropies), see Corollary \ref{g-c-c-c-1} below. \smallskip

It seems reasonable to conjecture that $"\Leftrightarrow"$ holds in (\ref{g-c-c-r}). This would give possibility to prove
preserving local continuity of the squashed entanglement under local operations, see the Conjecture after Corollary \ref{g-c-c-c-1}.\smallskip

Note also that the condition $\,\min\{H(\omega_{A}),H(\omega_{B})\}<+\infty\,$ in A is used only to show that
$E_{sq}(\omega^0_{AB})=\widehat{E}_{sq}(\omega^0_{AB})$ and $\,I(A\!:\!B)_{\omega^0}<+\infty$. To prove that $\mathrm{(iv)\Rightarrow(i)\Leftrightarrow(ii)\Leftrightarrow(iii)}$ it suffices to require  $\,I(A\!:\!B)_{\omega^0}<+\infty$. \smallskip
\end{remark}

\begin{remark}\label{g-c-c-rmk-2} Proposition \ref{g-c-c}C can be treated as a dominated convergence theorem for the squashed entanglement (cf.\cite{Simon}). In contrast to condition (\ref{H-c-c}) and Proposition \ref{g-c-c}A  it contains no assumptions concerning marginal entropies. Note that
in this case \emph{we do not assert} that property $\mathrm{(v)}$ holds (until it is not proved that $\,E_{sq}(\omega^0_{AB})=\widehat{E}_{sq}(\omega^0_{AB})$).
\end{remark}\smallskip

The proof of Proposition \ref{g-c-c} is based on the following two lemmas.\smallskip

\begin{lemma}\label{SE-n-c} \emph{Let $\shs\{\omega^k_{AB}\}$ be a sequence  converging to a state $\omega^0_{AB}$ and $\shs n\!\in\mathbb{N}$}\smallskip

A) \emph{The function $E^n_{sq}$ is lower semicontinuous on $\,\S(\H_{AB})$. If $\,I(A\!:\!B)_{\omega^0}\,$ is finite then}
\begin{equation*}
\liminf_{k\rightarrow\infty}E^n_{sq}(\omega_{AB}^k)-E^n_{sq}(\omega_{AB}^0)\geq
\liminf_{k\rightarrow\infty} I(A\!:\!B)_{\omega^k}-I(A\!:\!B)_{\omega^0}.
\end{equation*}

B) \emph{Local continuity of $\,I(A\!:\!B)$ is equivalent to local continuity of $E^n_{sq}$, i.e.}
\begin{equation*}%\label{SE-n-c-r}
\lim_{k\rightarrow\infty} I(A\!:\!B)_{\omega^k}=I(A\!:\!B)_{\omega^0}<+\infty\quad\Leftrightarrow\quad \lim_{k\rightarrow\infty}E^n_{sq}(\omega^k_{AB})=E^n_{sq}(\omega^0_{AB})<+\infty
\end{equation*}
\end{lemma}\medskip

\emph{Proof.} We may assume that for any state $\omega_{AB}$ the infimum in definition (\ref{se-def-n}) of $E^n_{sq}(\omega_{AB})$ is over all extensions $\omega_{ABE}$ in $\S(\H_{AB}\otimes\H^n_E)$, where $\H^n_E$ is a fixed $n$-dimensional Hilbert space.  By Lemma \ref{comp} the set of all such extensions of a given state $\omega_{AB}$ is compact. This and the lower semicontinuity of the function $\,\omega_{ABE}\mapsto I(A\!:\!B|E)_{\omega}$ (\cite[Th.2]{CMI}) imply attainability of the  infimum in definition (\ref{se-def-n}) of $E^n_{sq}(\omega_{AB})$.

For each $k$ let $\widetilde{\omega}^k_{ABE}\in\S(\H_{AB}\otimes\H^n_E)$ be an extension of the state $\omega^k_{AB}$ such that $E^n_{sq}(\omega_{AB}^k)=I(A\!:\!B|E)_{\widetilde{\omega}^k}$. By Lemma \ref{comp} the sequence $\tilde{\omega}^k_{ABE}$ is relatively compact and hence it has a limit point $\widetilde{\omega}^0_{ABE}\in\S(\H_{AB}\otimes\H^n_E)$. By continuity of a partial trace $\widetilde{\omega}^0_{ABE}$ is an extension of the state
$\omega^0_{AB}$  and hence $\,E^n_{sq}(\omega_{AB}^0)\leq I(A\!:\!B|E)_{\widetilde{\omega}^0}$.

Assume that $I(A\!:\!B)_{\omega^0}<+\infty$ and hence $E^n_{sq}(\omega_{AB}^0)<+\infty$. Let $\epsilon>0$ be arbitrary and $\{\omega^{k_t}_{AB}\}$ be a subsequence such that
$$
\lim_{t\rightarrow\infty} E^n_{sq}(\omega^{k_t}_{AB})\leq\liminf_{k\rightarrow\infty}E^n_{sq}(\omega_{AB}^k)+\epsilon\quad \textrm{and}\quad \lim_{t\rightarrow\infty}\tilde{\omega}^{k_t}_{ABE}=\tilde{\omega}^0_{ABE}.
$$
Since the functions $\,\omega_{ABE}\mapsto I(X\!:\!E)_{\omega}$, $X=A,B,AB$, are continuous on $\S(\H_{AB}\otimes\H^n_E)$, formula (\ref{cmi-d+++}) shows that
\begin{equation*}
\begin{array}{rl}
\displaystyle\liminf_{k\rightarrow\infty} I(A\!:\!B)_{\omega^k}-I(A\!:\!B)_{\omega^0}\leq
\displaystyle\liminf_{t\rightarrow\infty} I(A\!:\!B)_{\omega^{k_t}}-I(A\!:\!B)_{\omega^0}\\\\
=\displaystyle\lim_{t\rightarrow\infty} I(A\!:\!B|E)_{\widetilde{\omega}^{k_t}}-I(A\!:\!B|E)_{\widetilde{\omega}^0}
\leq\lim_{t\rightarrow\infty} E^n_{sq}(\omega^{k_t}_{AB})-E^n_{sq}(\omega_{AB}^0).
\end{array}
\end{equation*}
So, by the lower semicontinuity of $\,I(A\!:\!B)$, to prove the lower semicontinuity of  $E^n_{sq}$ on $\S(\H_{AB})$ it suffices to show that
$\,\displaystyle\lim_{k\rightarrow\infty} E^n_{sq}(\omega^k_{AB})=+\infty$ in the case $\,I(A\!:\!B)_{\omega^0}=+\infty$. This can be easily done by using formula (\ref{cmi-d+++}) and upper bound (\ref{mi-u-b}), since in this case the lower semicontinuity of $\,I(A\!:\!B)$ implies $\displaystyle\lim_{k\rightarrow\infty} I(A\!:\!B)_{\omega^k}=+\infty$. \smallskip

B) It suffices to show that local continuity of $\,I(A\!:\!B)$ implies local upper semicontinuity of $E^n_{sq}$.
Assume there exists a sequence $\{\omega^k_{AB}\}$ converging to a state $\omega^0_{AB}$ such that
\begin{equation}\label{c-assump}
\lim_{k\rightarrow\infty} I(A\!:\!B)_{\omega^k}=I(A\!:\!B)_{\omega}<+\infty\quad\textrm{and}\quad \lim_{k\rightarrow\infty}E^n_{sq}(\omega^k_{AB})> E^n_{sq}(\omega^0_{AB}).
\end{equation}
Let $\widetilde{\omega}^0_{ABE}\in\S(\H_{AB}\otimes\H^n_E)$ be an extension of the state
$\omega^0_{AB}$ such that $\,E^n_{sq}(\omega_{AB}^0)=I(A\!:\!B|E)_{\widetilde{\omega}^0}$. By using Lemma \ref{p-lemma} it is easy to show existence of a sequence
$\{\tilde{\omega}^k_{ABE}\}\subset\S(\H_{AB}\otimes\H^n_E)$  converging to the state
$\widetilde{\omega}^0_{ABE}$ such that $\widetilde{\omega}^k_{AB}=\omega^k_{AB}$ for all $k$.

Since the functions $\,\omega_{ABE}\mapsto I(X\!:\!E)_{\omega}$, $X=A,B,AB$, are continuous on $\S(\H_{AB}\otimes\H^n_E)$, formula (\ref{cmi-d+++}) and the first relation in (\ref{c-assump}) show that
$$
\lim_{k\rightarrow\infty} I(A\!:\!B|E)_{\widetilde{\omega}^k}=I(A\!:\!B|E)_{\widetilde{\omega}^0}=E^n_{sq}(\omega_{AB}^0).
$$
Since $\,E^n_{sq}(\omega_{AB}^k)\leq I(A\!:\!B|E)_{\widetilde{\omega}^k}$ for all $k$, this contradicts to the second relation in
(\ref{c-assump}). $\square$\smallskip

Lemmas \ref{sq-l} and \ref{SE-n-c}B imply the following observation. \smallskip

\begin{lemma}\label{SE-s-c} \emph{Local continuity of $\,I(A\!:\!B)$ implies local upper semicontinuity of $E_{sq}$, i.e. for any sequence $\{\omega^k_{AB}\}$  converging to a state $\omega^0_{AB}$ we have}
\begin{equation*}
\lim_{k\rightarrow\infty} I(A\!:\!B)_{\omega^k}=I(A\!:\!B)_{\omega^0}<+\infty\;\;\Rightarrow\;\;\limsup_{k\rightarrow\infty}E_{sq}(\omega^k_{AB})\leq E_{sq}(\omega^0_{AB})<+\infty.
\end{equation*}
\end{lemma}\medskip
\emph{Proof of  Proposition \ref{g-c-c}.} A) The condition $\,\min\{H(\omega^0_{A}),H(\omega^0_{B})\}<+\infty$ implies, by Propositions \ref{two-def}, that
\begin{equation}\label{E-ineq}
\widehat{E}_{sq}(\omega^0_{AB})=E_{sq}(\omega^0_{AB})\leq I(A\!:\!B)_{\omega^0}<+\infty.
\end{equation}

Since $I(A\!:\!B)_{\omega^0}$ is finite, $\mathrm{(i)\Leftrightarrow(ii)\Leftrightarrow(iii)}$ follows from Lemma \ref{SE-n-c}.\smallskip

$\mathrm{(i)\Rightarrow(iv)\Rightarrow(v)}$. Lemma \ref{SE-s-c} shows that
\begin{equation}\label{E-usc}
 \limsup_{k\rightarrow\infty}E_{sq}(\omega^k_{AB})\leq E_{sq}(\omega^0_{AB}).
\end{equation}
Since $\widehat{E}_{sq}$ is lower semicontinuous on $\S(\H_{AB})$, we have
\begin{equation}\label{E-lsc}
 \liminf_{k\rightarrow\infty}\widehat{E}_{sq}(\omega^k_{AB})\geq \widehat{E}_{sq}(\omega^0_{AB}).
\end{equation}
Since $\widehat{E}_{sq}(\omega^k_{AB})\leq E_{sq}(\omega^k_{AB})$ for all $k$,
(\ref{E-ineq}),(\ref{E-usc}) and (\ref{E-lsc}) imply
$$
\lim_{k\rightarrow\infty}E_{sq}(\omega^k_{AB})=E_{sq}(\omega^0_{AB})=\widehat{E}_{sq}(\omega^0_{AB})=
\lim_{k\rightarrow\infty}\widehat{E}_{sq}(\omega^k_{AB}).
$$

It follows from $\mathrm{(i)}$ that $I(A\!:\!B)_{\omega^k}<+\infty$ for all $k\geq k_*$.  Lemma \ref{sq-l} and $\,\mathrm{(i)\Leftrightarrow(iii)}\,$  show that the monotone sequence $\{E^n_{sq}\}$ of continuous functions pointwise converges to the continuous function $E_{sq}$ on the compact set $\{\omega_{AB}^0,\omega_{AB}^{k_*},\omega_{AB}^{k_*+1},...\}$. By Dini's lemma the sequence $\{E^n_{sq}\}$ converges uniformly on this set implying the second relation in $\mathrm{(iv)}$.

$\mathrm{(iv)\Rightarrow(i)}$. Assume that $\mathrm{(i)}$ is not valid. By lower semicontinuity of $I(A\!:\!B)$ we may consider (by passing to a subsequence) that
$$
\liminf_{k\rightarrow\infty} I(A\!:\!B)_{\omega^k}\geq I(A\!:\!B)_{\omega^0}+\Delta
$$
for some $\Delta>0$. So, Lemma \ref{SE-n-c}A implies
\begin{equation*}
  \liminf_{k\rightarrow\infty}E^n_{sq}(\omega_{AB}^k)\geq E^n_{sq}(\omega_{AB}^0)+\Delta \;\;\textrm{ for all }\; n.
\end{equation*}
Since $E^n_{sq}(\omega_{AB}^0)$ tends to $E_{sq}(\omega_{AB}^0)$ by Lemma \ref{sq-l}, this contradicts to  $\mathrm{(iv)}$. \medskip

B) Since in this case $\,I(A\!:\!B)_{\omega^0}=+\infty$,  inequality (\ref{u-ineq})
and Proposition \ref{two-def} imply $\,E_{sq}(\omega_{AB}^0)=\widehat{E}_{sq}(\omega_{AB}^0)=+\infty$. So, to prove this assertion it suffices to note
that $\,\widehat{E}_{sq}\leq E_{sq}\leq E^n_{sq}\leq I(A\!:\!B)\,$ for all $\,n\,$ and to use the lower semicontinuity of $\,\widehat{E}_{sq}$.\medskip

C) The conditions imply (cf.\cite[Th.1]{CMI})
\begin{equation}\label{d-ineq}
\lim_{k\rightarrow\infty}I(A\!:\!B)_{\omega^k}=I(A\!:\!B)_{\omega^0}<+\infty.
\end{equation}
By the convexity of $E_{sq}$ we have
$$
E_{sq}(\omega^0_{AB})\leq \lambda_k E_{sq}(\omega^k_{AB})+(1-\lambda_k)E_{sq}(\widetilde{\omega}^k_{AB})\leq \lambda_k E_{sq}(\omega^k_{AB})+(1-\lambda_k)I(A\!:\!B)_{\widetilde{\omega}^k},
$$
where $\widetilde{\omega}^k_{AB}=(1-\lambda_k)^{-1}(\omega^0_{AB}-\lambda_k\omega^k_{AB})$ is bona fide state for each $k$. So, to prove that
$\lim_{k\rightarrow\infty}E_{sq}(\omega^k_{AB})=E_{sq}(\omega^0_{AB})$ it suffices to show, by Lemma \ref{SE-s-c}, that
\begin{equation*}
 \lim_{k\rightarrow\infty}(1-\lambda_k)I(A\!:\!B)_{\widetilde{\omega}^k}=0
\end{equation*}
This follows from (\ref{d-ineq}), since inequality (\ref{mi-conc}) implies
$$
(1-\lambda_k)I(A\!:\!B)_{\widetilde{\omega}^k}\leq I(A\!:\!B)_{\omega^0}-\lambda_kI(A\!:\!B)_{\omega^k}+2h_2(\lambda_k).
$$

By Lemma \ref{SE-n-c} relation (\ref{d-ineq}) implies $\mathrm{(iii)}$. So, the second part of $\mathrm{(iv)}$ follows from Lemma \ref{sq-l} and Dini's lemma. $\square$ \smallskip

By Theorem 1B in \cite{CMI} local continuity of quantum mutual information is preserved by local operations, i.e. for any sequence $\{\omega^k_{AB}\}$ of states converging to a state $\omega^0_{AB}$ and arbitrary quantum operations $\Phi_A:A\rightarrow A$ and $\Phi_B:B\rightarrow B$ we have
\begin{equation*}
\lim_{k\rightarrow\infty} I(A\!:\!B)_{\omega^k}=I(A\!:\!B)_{\omega^0}<+\infty\;\;\Rightarrow\;\;\lim_{k\rightarrow\infty} I(A\!:\!B)_{\widetilde{\omega}^k}=I(A\!:\!B)_{\widetilde{\omega}^0}<+\infty,
\end{equation*}
where $\widetilde{\omega}^k_{AB}=\Phi_A\otimes\Phi_B(\omega^k_{AB})$ for all $k$.\footnote{If $\Phi_A$ and $\Phi_B$ are not channels then $I(A\!:\!B)_{\tilde{\omega}^k}$ in the above relation is defined by formula (\ref{mi-ext}).}  This implication, Proposition \ref{g-c-c}A and the last sentence in Remark \ref{g-c-c-rmk-1} imply the
following observation.\smallskip

\begin{corollary}\label{g-c-c-c-1} \emph{If one of properties $\mathrm{(i)\textrm{-}(iv)}$ in Proposition \ref{g-c-c}A holds for a sequence $\,\{\omega^k_{AB}\}$  converging to a state $\,\omega^0_{AB}$ such that $I(A\!:\!B)_{\omega^0}<+\infty$ then all the properties $\mathrm{(i)\textrm{-}(v)}$ hold for the sequence  $\{\widetilde{\omega}^k_{AB}\}$, where
$$
\widetilde{\omega}^k_{AB}=[\Tr\shs\Phi_A\otimes\Phi_B(\omega^k_{AB})]^{-1}\Phi_A\otimes\Phi_B(\omega^k_{AB}),
$$
for any local quantum operations
$\,\Phi_A:A\rightarrow A$ and $\,\Phi_B:B\rightarrow B$ such that $\Phi_A\otimes\Phi_B(\omega^0_{AB})\neq0$ and $\,\min\left\{H(\Phi_A(\omega^0_{A})),H(\Phi_B(\omega^0_{B}))\right\}<+\infty$. In particular,
\begin{equation*}
\lim_{k\rightarrow\infty} I(A\!:\!B)_{\omega^k}=I(A\!:\!B)_{\omega^0}<+\infty\quad\Rightarrow\quad \lim_{k\rightarrow\infty} E(\widetilde{\omega}^k_{AB})=E(\widetilde{\omega}^0_{AB})<+\infty,
\end{equation*}
for $E=E_{sq},\widehat{E}_{sq}$.}\smallskip
\end{corollary}\medskip

The main assertion of Corollary \ref{g-c-c-c-1} is a weak form of the following\smallskip

\textbf{Conjecture:} \emph{Local continuity of the squashed entanglement is preserved by local operations}.
\smallskip

To prove this conjecture it suffices to show that  $"\Leftrightarrow"$ holds in (\ref{g-c-c-r}).
\smallskip

Theorem 1A in \cite{CMI} and Proposition \ref{g-c-c}A imply the following continuity condition. \smallskip
\begin{corollary}\label{g-c-c-c-2} \emph{Let $\omega^0_{AB}$ be a state such that $\min\left\{H(\omega^0_{A}),H(\omega^0_{B})\right\}<+\infty$
and $\{\omega_{AB}^k\}$ be a sequence of states converging to the state $\omega^0_{AB}$ such that $\lambda_k\omega^k\leq\Phi_A^k\otimes\Phi_B^k(\omega^0)$ for some local quantum operations $\Phi_A^k$ and $\Phi_B^k$, where
$\,\{\lambda_k\}$ is a sequence converging to $1$. Then}
\begin{equation}\label{E-l-r}
\lim_{k\rightarrow\infty} E_{sq}(\omega^k_{AB})=\lim_{k\rightarrow\infty}\widehat{E}_{sq}(\omega^k_{AB})=E_{sq}(\omega^0_{AB})=\widehat{E}_{sq}(\omega^0_{AB})<+\infty.
\end{equation}
\end{corollary}\medskip

Corollary \ref{g-c-c-c-2} shows in particular that (\ref{E-l-r}) holds if $\{\omega^k_{AB}\}$ is a sequence of states proportional to the operators
$$
\Tr_C\Phi_A^k\otimes\Phi_B^k\otimes\Phi_C^k(\omega_{ABC}^0),
$$
where $\omega^0_{ABC}$ is a state such that $\min\left\{H(\omega^0_{A}),H(\omega^0_{B})\right\}<+\infty$ and $\{\Phi_A^k\}$,$\{\Phi_B^k\}$, $\{\Phi_C^k\}$ are sequences of local quantum operations strongly converging to the identity channels $\id_A$,$\id_B$,$\id_C$ (this means that $\lim_{k}\Phi_X^k(\rho)=\rho$ for arbitrary states $\rho\in\S(\H_X)$, $X=A,B,C$ \cite{AQC}).\smallskip

\begin{corollary}\label{se-pr-c} \emph{If one of the systems $A$ and $B$, say $A$, is finite-dimensional then
$E_{sq}=\widehat{E}_{sq}=E^*_{sq}$ (the function $E^*_{sq}$ is defined by (\ref{se-def-s})) is a continuous entanglement measure on $\,\S(\H_{AB})$ and\vspace{-5pt}
$$
\left|E_{sq}(\omega_{AB}^2)-E_{sq}(\omega_{AB}^1)\right|\leq \sqrt{\varepsilon}
\log\dim\H_A+2(1+\sqrt{\varepsilon})h_2\!\left(\frac{\sqrt{\varepsilon}}{1+\sqrt{\varepsilon}}\right),
$$\vspace{-5pt}
for any $\;\omega_{AB}^1$, $\,\omega_{AB}^2\,$ such that $\;\varepsilon=\|\omega_{AB}^2-\omega_{AB}^1\|_1<1$.}
\end{corollary}
\medskip

\emph{Proof.} Lemma \ref{sq-l} and Proposition \ref{two-def} imply $E_{sq}=\widehat{E}_{sq}=E^*_{sq}$. To prove the continuity bound it suffices to note that in this case $E_{sq}$ coincides with the function $f_n$ in the proof of assertion C of Proposition \ref{se-pr} for some finite $n$ and to repeat the arguments from that proof. $\square$

\subsection{Continuity bounds for $E_{sq}$ and for $E_{F}$ under the energy constraint on one subsystem}

In this subsection we restrict attention to subsets of bipartite states $\,\omega_{AB}$ with bounded energy of
$\omega_{A}$, i.e. subsets of the form
\begin{equation}\label{s-b-e}
\S_E\doteq\{\shs\omega_{AB}\,|\,\Tr H_A\omega_{A}\leq E\shs\},
\end{equation}
where $H_A$ is a Hamiltonian of the system $A$ and $E>0$.

It is well known that the von Neumann entropy is bounded on the set $\,\{\rho_A\,|\,\Tr H_A\rho_A\leq E\}$ if and only if $\,\Tr e^{-\beta H_A}<+\infty$  for some $\beta>0$ and it is continuous on this set if and only if $\,\Tr e^{-\beta H_A}<+\infty$ for all $\beta>0$ \cite{EC,W}. So, Propositions \ref{se-pr}D and \ref{two-def}C imply the following assertions. \smallskip

\begin{corollary}\label{two-def-c+} A) \emph{If
$\,\Tr e^{-\beta H_A}$  is finite for some $\,\beta>0$ then the functions $E_{sq}$ and $\widehat{E}_{sq}$ coincide and are lower semicontinuous on  $\,\S_E$ for any $\,E>0$.} \smallskip

B) \emph{If $\,\Tr e^{-\beta H_A}$ is finite for all $\,\beta>0$ then the functions $E_{sq}$ and $\widehat{E}_{sq}$ coincide and are continuous on  $\,\S_E$ for any $\,E>0$.}\smallskip

C) \emph{Assertions A and B also hold for the infinite-dimensional versions $E_F^{d}$ and  $E_F^{c}$ of the entanglement of formation (considered at the end of Sect.3).}
\end{corollary}\medskip

The last assertion of Corollary \ref{two-def-c+} follows from the results in \cite[Sect.5]{EM}.
\smallskip

Corollary \ref{two-def-c+} implies coincidence and continuity of the functions $E_{sq}$ and $\widehat{E}_{sq}$ (and of the functions $E_F^{d}$ and  $E_F^{c}$) on the set of states of a bipartite finite-mode Bosonic system with bounded mean energy \cite[Ch.12]{H-SCI}.\medskip

Corollary \ref{two-def-c+} can be strengthened by using the approach recently proposed by Winter in \cite{Winter}, where the tight continuity bounds for the von Neumann entropy and for the conditional entropy under the energy constraint are obtained. Combining this approach, Lemma \ref{main-l+} in Section 4 and the Fannes type continuity bound for the conditional mutual information one can obtain  tight continuity bound for the conditional mutual information under the energy constraint on one subsystem (Lemma \ref{c-b-cmi} in the Appendix), which makes possible to derive continuity bounds for the squashed entanglement and for the entanglement of formation under the same constraint.

Let $H_A$ be a Hamiltonian of system $A$ such that
\begin{equation}\label{g-c}
 Z(\beta)\doteq\Tr e^{-\beta H_A}<+\infty\quad\textrm{for all}\quad\beta>0.
\end{equation}
Then for any $E>0$ the von Neumann entropy attains its maximum on the set $\,\{\rho_A\,|\,\Tr H_A\rho_A\leq E\}$ at the Gibbs state $\gamma(E)=[Z(\beta(E))]^{-1} e^{-\beta(E) H_A}$, where $\beta(E)$ is the solution of the equation $\Tr H_Ae^{-\beta H_A}=EZ(\beta)$ \cite{W}. Condition (\ref{g-c}) implies that $H_A$ has discrete spectrum $\{E_1, E_2,...\}$ of finite multiplicity. We will assume  that $E_1=0$.\smallskip

We will obtain continuity bound for the squashed entanglement on the set $\S_E$ defined by (\ref{s-b-e}) and will show that \emph{the same}  continuity bound is valid for the entanglement of formation. Corollary \ref{two-def-c+} shows that  $\,E_{sq}(\omega_{AB})=\widehat{E}_{sq}(\omega_{AB})\,$ and $\,E^d_{F}(\omega_{AB})=E^c_{F}(\omega_{AB})\,$ for any state $\,\omega_{AB}\in\S_E$. So,  in what follows we will forget about possible noncoincidence of $E_{sq}$ with $\widehat{E}_{sq}$ and of $E^d_{F}$ with $E^c_{F}$ and will denote these functions respectively by  $E_{sq}$ and by $E_{F}$.\smallskip

\begin{property}\label{c-b-em} \emph{Let $\,\omega^1_{AB}\,$ and $\,\omega^2_{AB}\,$ be arbitrary states in $\S_E$ such that   $\|\omega^2_{AB}-\omega^1_{AB}\|_1=\varepsilon<1$. Let $\,\varepsilon'\in(\sqrt{\varepsilon},1]$ and $\,\delta=\frac{\varepsilon'-\sqrt{\varepsilon}}{1+\varepsilon'}$. Then
$$
\left|E_{sq}(\omega^2_{AB})-E_{sq}(\omega^1_{AB})\right|\leq(\varepsilon'+2\delta)H\!\left(\gamma(E/\delta)\right)
+2(1+\varepsilon')h_2\!\left(\!\frac{\varepsilon'}{1+\varepsilon'}\!\right)+2h_2(\delta)
$$
The same continuity bound is valid for the entanglement of formation $E_F$.}
\end{property}\medskip

The coincidence of the continuity bounds for $E_{sq}$ and for $E_F$ is not surprising due to representation (\ref{ef-ref}).
\medskip

\begin{remark}\label{c-b-em-r} By Proposition 1 in \cite{EC} condition (\ref{g-c}) implies
 \begin{equation}\label{main-lr}
   \lim_{\delta\rightarrow+0}\delta H(\gamma(E/\delta))=0.
 \end{equation}
 So, the continuity bounds in Proposition \ref{c-b-em} show uniform continuity of  the squashed entanglement and of the entanglement of formation on the set $\S_E$.
\end{remark}

\smallskip

\emph{Proof.} By repeating
the arguments from the proof of continuity of $E_{sq}$ in \cite{C&W} we obtain
\begin{equation}\label{sq-rep+}
E_{sq}(\omega^k_{AB})=\textstyle\frac{1}{2}\displaystyle\inf_{\Lambda}I(A\!:\!B|E)_{\id_{AB}\otimes\Lambda(\omega^k_{ABC})},\quad k=1,2,
\end{equation}
where $\omega^1_{ABC}$  and $\omega^2_{ABC}$ are purifications of
$\omega^1_{AB}$ and of $\omega^2_{AB}$ such that
$$\|\omega^2_{ABC}-\omega^1_{ABC}\|_1\leq2\sqrt{\varepsilon},$$
and the infimum is over all quantum channels
$\Lambda:\T(\H_C)\rightarrow\T(\H_E)$.\smallskip

For given $\Lambda$ let
$\omega^1_{ABE}=\id_{AB}\otimes\Lambda(\omega^1_{ABC})$ and
$\omega^2_{ABE}=\id_{AB}\otimes\Lambda(\omega^2_{ABC})$. Then
$\|\omega^2_{ABE}-\omega^1_{ABE}\|_1\leq2\sqrt{\varepsilon}$,
and Lemma \ref{c-b-cmi} in the Appendix implies
$$
\left|I(A\!:\!B|E)_{\omega^2}-I(A\!:\!B|E)_{\omega^1}\right|\leq (2\varepsilon'+4\delta)H\!\left(\gamma(E/\delta)\right)
+4\theta(\varepsilon')+4h_2(\delta),
$$
where $\theta(x)=(1+x)h_2\!\left(\frac{x}{1+x}\right)$. Since this estimate holds for any quantum channel $\Lambda$, expression (\ref{sq-rep+}) implies the above continuity bound for $E_{sq}$. \smallskip

To obtain the continuity bound for $E_{F}$ we will use the modification of Nielsen's technique \cite{Nielsen}. Let $\epsilon>0$ be arbitrary and
$\{\pi_i,\varpi^i_{AB} \}$ be an ensemble (finite or countable) of pure states such that
$$
\omega^1_{AB}=\sum_i\pi_i\varpi^i_{AB}\quad \textrm{and} \quad E_{F}(\omega^1_{AB})\geq \sum_i\pi_iH(\varpi^i_A)-\epsilon.
$$
Let $R$ be an infinite-dimensional reference system,  $\omega^1_{ABR}$ and $\omega^2_{ABR}$ be purifications of $\omega^1_{AB}$ and of $\omega^2_{AB}$ such that
$$\|\omega^2_{ABR}-\omega^1_{ABR}\|_1\leq2\sqrt{\varepsilon}.
$$
By the arguments from \cite{W&Co} we can choose such basic $\{|i\rangle\}$ in $\H_R$ that
$$
\tilde{\omega}_{ABR}^1=\sum_iI_{AB}\otimes|i\rangle\langle i|\,\omega^1_{ABR}\,I_{AB}\otimes|i\rangle\langle i|=\sum_i\pi_i\varpi^i_{AB}\otimes|i\rangle\langle i|.
$$
Thus,
$$
\sum_i\pi_iH(\varpi^i_A)=\textstyle\frac{1}{2}\shs I(A:B|R)_{\tilde{\omega}^1}\;\;\text{and hence}\;\; E_{F}(\omega^1_{AB})\geq \textstyle\frac{1}{2}\shs I(A:B|R)_{\tilde{\omega}^1}-\epsilon.
$$

On the other hand, since
$$
\tilde{\omega}_{ABR}^2=\sum_iI_{AB}\otimes|i\rangle\langle i|\,\omega^2_{ABR}\,I_{AB}\otimes|i\rangle\langle i|=\sum_i|\psi_i\rangle\langle\psi_i|\otimes|i\rangle\langle i|,
$$
where $\{|\psi_i\rangle\}$ is a collection of vectors such that $\,\sum_i|\psi_i\rangle\langle\psi_i|=\tilde{\omega}_{AB}^2=\omega_{AB}^2$,
it is easy to see that $\,\frac{1}{2}\shs I(A:B|R)_{\tilde{\omega}^2}\geq E_{F}(\omega^2_{AB})$.

So, we have
$$
E_{F}(\omega^2_{AB})-E_{F}(\omega^1_{AB})\leq \textstyle\frac{1}{2}\shs I(A:B|R)_{\tilde{\omega}^2}-\textstyle\frac{1}{2}\shs I(A:B|R)_{\tilde{\omega}^1}+\epsilon.
$$
Since $\|\tilde{\omega}^2_{ABR}-\tilde{\omega}^1_{ABR}\|_1\leq2\sqrt{\varepsilon}$, by applying Lemma \ref{c-b-cmi} in the Appendix we obtain
$$
E_{F}(\omega^2_{AB})-E_{F}(\omega^1_{AB})\leq(\varepsilon'+2\delta)H\!\left(\gamma(E/\delta)\right)
+2(1+\varepsilon')h_2\!\left(\!\frac{\varepsilon'}{1+\varepsilon'}\!\right)+2h_2(\delta)+\epsilon.
$$
By permuting $\omega^2_{AB}$ and $\omega^1_{AB}$ in the above argumentation we obtain the same upper bound for
$E_{F}(\omega^1_{AB})-E_{F}(\omega^2_{AB})$. Since $\,\epsilon\,$ is arbitrary, this gives the  continuity bound for $E_F$ coinciding with the continuity bound for $E_{sq}$. $\square$\medskip

Proposition \ref{c-b-em} implies the following asymptotic continuity property of the squashed  entanglement and of
the entanglement of formation (cf. \cite{ESP}). \smallskip

\begin{corollary}\label{c-b-em-c} \emph{Let $H_{A^n}=H_A\otimes I_A\otimes\ldots\otimes I_A+\ldots+I_A\otimes\ldots\otimes I_A\otimes H_A$
be a Hamiltonian of the system $A^{n}$ and
$\,\{\omega^1_{n}\},$ $\,\{\omega^2_{n}\}\,$ be sequences of states such that
$$
\omega^k_{n}\in\S(\H^{\otimes n}_{AB}),\quad \Tr H_{A^n}[\omega^k_{n}]_{A^n}\leq nE,\quad k=1,2, \qquad \lim_{n\rightarrow\infty}\|\omega^2_{n}-\omega^1_{n}\|_1=0.
$$
If $\,H_A$ satisfies condition (\ref{g-c}) then
$$
\lim_{n\rightarrow\infty}\frac{\left|E_{sq}(\omega^2_{n})-E_{sq}(\omega^1_{n})\right|}{n}=0\quad \textit{and}\quad
\lim_{n\rightarrow\infty}\frac{\left|E_{F}(\omega^2_{n})-E_{F}(\omega^1_{n})\right|}{n}=0.
$$}
\end{corollary}\medskip

\emph{Proof.} Note that $H_{A^n}$ satisfies condition (\ref{g-c}) and that $\gamma(E)^{\otimes n}$
is the Gibbs state of the system $A^n$ corresponding to the energy $nE$. So, the above limit relations directly follow from the continuity bounds in Proposition \ref{c-b-em} and relation (\ref{main-lr}). $\square$

\section{Conclusion}

We have presented a detailed analysis of properties of the squashed entanglement $E_{sq}$ obtained by direct translation
(\ref{se-def+}) of the finite-dimensional definition  in comparison with the universal extension $\widehat{E}_{sq}$ defined by formula (\ref{Esq-def}). The function $\widehat{E}_{sq}$ is a result of application to the case of squashed entanglement of the general method for  construction of infinite-dimensional entanglement monotones starting from finite-dimensional ones described in Section 3.

It is shown that the functions $E_{sq}$ and $\widehat{E}_{sq}$  produce the same  entanglement measure on the set
$$
\S_\mathrm{*}\doteq\left\{\shs\omega_{AB}\,|\,\min\{H(\omega_{A}),H(\omega_{B}),H(\omega_{AB})\}<+\infty\shs\right\}
$$
possessing all basis properties of the squashed entanglement valid in finite dimensions (with continuity replaced by lower semicontinuity). It is proved that the function $\widehat{E}_{sq}$  is an adequate lower semicontinuous extension of  this measure to the set $\S(\H_{AB})$ of all bipartite states. Coincidence of $E_{sq}$ and $\widehat{E}_{sq}$ on $\S(\H_{AB})$ is a conjecture  equivalent to  lower semicontinuity of $E_{sq}$ on $\S(\H_{AB})$.

It is shown that local continuity  of the quantum mutual information  implies local continuity of the squashed entanglement (Proposition \ref{g-c-c}, Remark \ref{g-c-c-rmk-1}). A weak form of the conjecture that local continuity of the squashed entanglement is preserved  by local operations is proved (Corollary \ref{g-c-c-c-1}).

The common continuity bound for the squashed entanglement and for the entanglement of formation under the energy constraint on one subsystem is obtained and used to prove the asymptotic continuity of these entanglement measures under the same constraint (Proposition \ref{c-b-em}, Corollary \ref{c-b-em-c}).

Some open questions are formulated. One of them concerns possibility to extend a countably nondecomposable separable state to a short Markov chain (Remark \ref{c-n-s}).

\section{Appendix: Tight continuity bound for conditional mutual information under the energy constraint on one subsystem}

Since for any state $\omega_{ABC}$ with finite $H(\omega_A)$ we have
$$
I(A\!:\!C|B)_{\omega}=H(A|B)_{\omega}-H(A|BC)_{\omega},
$$
where $H(A|X)$ is the extended conditional entropy defined in (\ref{c-e-d+}), continuity bound for
$I(A\!:\!C|B)_{\omega}$ under the energy constraint on $\omega_A$ can be directly obtained from Meta-Lemma 14 in \cite{Winter} (by proving that this lemma remans valid for the extended conditional entropy).

But more sharp continuity bound for $I(A\!:\!C|B)_{\omega}$ can be obtained by using Winter's technique (rather than his final results) and some estimates for conditional mutual information.

Let $H_A$ be a Hamiltonian of system A such that $\,Z(\beta)\doteq\Tr e^{-\beta H_A}$ is finite for all $\,\beta>0$.
This implies that $H_A$ has discrete spectrum of finite multiplicity, i.e. $H_A=\sum_{n=1}^{+\infty} E_n|n\rangle\langle n|$, where $\{|n\rangle\}_{n=1}^{+\infty}$ is an orthonormal basis of eigenvectors of $H_A$  corresponding to the nondecreasing sequence $\{E_n\}_{n=1}^{+\infty}$ of eigenvalues (energy levels of $H_A$) such that $\sum_{n=1}^{+\infty} e^{-\beta E_n}$ is finite for all $\,\beta>0$. We will assume that $E_1=0$ for simplicity.\smallskip

For any $E>0$ the von Neumann entropy attains its maximum under the constraint $\Tr \rho H_A\leq E$ at the Gibbs state $\gamma(E)=[Z(\beta(E))]^{-1}e^{-\beta(E) H_A}$, where $\beta(E)$ is the solution of the equation $\Tr H_Ae^{-\beta H_A}=EZ(\beta)$ \cite{W}. \smallskip

\begin{lemma}\label{c-b-cmi} \emph{Let $\,\rho\,$ and $\,\sigma\,$ be states of the tripartite system $ABC$ such that $\,\Tr \rho_A H_A,\Tr \sigma_A H_A\leq E$, $\frac{1}{2}\|\rho-\sigma\|_1\leq\varepsilon<\varepsilon'\leq1\,$ and $\,\delta=\frac{\varepsilon'-\varepsilon}{1+\varepsilon'}$. Then}
$$
\left|I(A\!:\!C|B)_{\rho}-I(A\!:\!C|B)_{\sigma}\right|\leq(2\varepsilon'+4\delta)H\!\left(\gamma(E/\delta)\right)
+4(1+\varepsilon')h_2\!\left(\!\frac{\varepsilon'}{1+\varepsilon'}\!\right)+4h_2(\delta).
$$
\end{lemma}

\begin{remark}\label{c-b-cmi-r-1} Since $\,\displaystyle \lim_{\delta\rightarrow+0}\delta H(\gamma(E/\delta))=0\,$ (see Remark \ref{c-b-em-r}), Lemma \ref{c-b-cmi} shows that the function $\omega_{ABC}\mapsto I(A\!:\!C|B)_{\omega}$ is uniformly continuous on the set $\{\omega_{ABC}\,|\,\Tr H_A\omega_A\leq E\}$ for any $E>0$.
\end{remark}

\medskip

\emph{Proof.} Following the proofs  of Lemmas 13,14 in \cite{Winter} define the projectors
$$
P_{\leq}\doteq\sum_{0\leq E_n\leq E/\delta}|n\rangle\langle n|
$$
and consider the states
$$
\rho_{\leq}=\frac{P_{\leq}\rho P_{\leq}}{\Tr P_{\leq}\rho},\quad \sigma_{\leq}=\frac{P_{\leq}\sigma P_{\leq}}{\Tr P_{\leq}\sigma}.
$$
In the proof of Lemma 13 in \cite{Winter} it is shown that
\begin{equation}\label{est-1}
  H(\omega)-[\Tr P_{\leq}\omega] H(\omega_{\leq})\leq \delta H\!\left(\gamma(E/\delta)\right)+h_2(\Tr P_{\leq}\omega),
\end{equation}
\begin{equation}\label{est-2}
H(\omega_{\leq})\leq  H\!\left(\gamma(E/\delta)\right),\quad\Tr P_{\leq}\omega\geq 1-\delta,
\end{equation}
where $\,\omega=\rho,\sigma$, and that
\begin{equation}\label{est-3}
\log\Tr P_{\leq}\leq H\!\left(\gamma(E/\delta)\right),\quad    \textstyle\frac{1}{2}\|\rho_{\leq}-\sigma_{\leq}\|\leq\varepsilon'.
\end{equation}
By using (\ref{est-1}) and (\ref{est-2}) it is easy to derive from the second inequality of Lemma \ref{main-l+} that
\begin{equation}\label{est-4}
\left|I(A\!:\!C|B)_{\omega}-I(A\!:\!C|B)_{\omega_{\leq}}\right|\leq2\delta H\!\left(\gamma(E/\delta)\right)+2h_2(\delta),\quad \omega=\rho,\sigma.
\end{equation}
By using (\ref{est-3}) and applying Corollary 8 in \cite{CMI} (Fannes type continuity bound for conditional mutual information)
we obtain
\begin{equation}\label{est-5}
\begin{array}{c}
\left|I(A\!:\!C|B)_{\rho_{\leq}}-I(A\!:\!C|B)_{\sigma_{\leq}}\right|\leq2\varepsilon'\log\Tr P_{\leq}+4(1+\varepsilon')h_2\!\left(\frac{\varepsilon'}{1+\varepsilon'}\right)\\\\\leq
2\varepsilon'H\!\left(\gamma(E/\delta)\right)+4(1+\varepsilon')h_2\!\left(\frac{\varepsilon'}{1+\varepsilon'}\right).
\end{array}
\end{equation}
Since
$$
\begin{array}{c}
\left|I(A\!:\!C|B)_{\rho}-I(A\!:\!C|B)_{\sigma}\right|\leq\left|I(A\!:\!C|B)_{\rho_{\leq}}-I(A\!:\!C|B)_{\sigma_{\leq}}\right|\\\\
\qquad+\left|I(A\!:\!C|B)_{\rho}-I(A\!:\!C|B)_{\rho_{\leq}}\right|+\left|I(A\!:\!C|B)_{\sigma}-I(A\!:\!C|B)_{\sigma_{\leq}}\right|,
\end{array}
$$
the required continuity bound follows from (\ref{est-4}) and (\ref{est-5}). $\square$
\medskip

\begin{remark}\label{c-b-cmi-r-2}
By using Lemma \ref{c-b-cmi} one can obtain analog of Lemma 15 in \cite{Winter} for conditional mutual information, i.e. a continuity bound for conditional mutual information $I(A\!:\!C|B)$ under the energy constraint on the system $A$ composed of $\ell$ oscillators (assuming that $B$ and $C$ are arbitrary systems).  Since the main terms in Lemma \ref{c-b-cmi} and in Meta-Lemma 14 in \cite{Winter} coincide,
the main term in this continuity bound coincides with the main term
$$
2\varepsilon\left(\frac{1+\alpha}{1-\alpha}+2\alpha\right)\sum_{i=1}^{\ell}\log\left(\frac{\bar{E}}{\hbar\omega_i}+1\right),\quad \bar{E}=E/\ell,
$$
in the continuity bound for conditional entropy in Lemma 15 in \cite{Winter}, where $\,\omega_1,...,\omega_\ell\,$ are frequencies of the oscillators and $\,\alpha\in(0,1/2)$ is a free parameter. By choosing arbitrarily small $\alpha$ and large energy $E$ we see (as in Remark 16 in \cite{Winter}) that this term is approximately equal to $2\varepsilon H\!\left(\gamma(E)\right)$. \smallskip

Let $\rho=\gamma(E)\otimes\tau_{B}\otimes\tau_C$ and
$\sigma=(1-\varepsilon)\rho+\varepsilon|\phi_{AC}\rangle\langle\phi_{AC}|\otimes\tau'_B$, where $\,|\phi_{AC}\rangle\langle\phi_{AC}|\in\S(\H_{AC})\,$ is a purification of the Gibbs state $\gamma(E)\in\S(\H_{A})$,  $\tau_B$ and $\tau'_B$ are orthogonal pure states in $\S(\H_B)$, $\tau_C$ is a pure state in $\S(\H_C)$. Then  $\,I(A\!:\!C|B)_{\rho}=0\,$ and  $I(A\!:\!C|B)_{\sigma}=2\varepsilon H\!\left(\gamma(E)\right)$, so that
$$
I(A\!:\!C|B)_{\sigma}-I(A\!:\!C|B)_{\rho}=2\varepsilon H\!\left(\gamma(E)\right).
$$
Since $\frac{1}{2}\|\sigma-\rho\|_1\leq\varepsilon$, we conclude that Lemma \ref{c-b-cmi} gives \emph{asymptotically tight}
continuity bound for conditional mutual information.
\end{remark}

\bigskip

I am grateful to A.S.Holevo and to the participants of his seminar
"Quantum probability, statistic, information" (the Steklov
Mathematical Institute) for useful discussion. I am also grateful to A.Winter for valuable communication.
The research is funded by the grant of Russian Science Foundation
(project No 14-21-00162).

\end{document}